\documentclass[12pt,preprint2]{aastex}
\usepackage{natbib,makeidx}
\usepackage{lscape}

\newcommand{\Ha}{\mbox{H$\alpha$}}

\newcommand{\etal}{\mbox{et al.}}

\newcommand{\vj}{\mbox{$V\!-\!J$}}
\newcommand{\vh}{\mbox{$V\!-\!H$}}
\newcommand{\vk}{\mbox{$V\!-\!K$}}
\newcommand{\bk}{\mbox{$B\!-\!K$}}
\newcommand{\jh}{\mbox{$J\!-\!H$}}
\newcommand{\hk}{\mbox{$H\!-\!K$}}
\newcommand{\jk}{\mbox{$J\!-\!K$}}
\newcommand{\ks}{\mbox{$K_s$}}

\newcommand{\sbb}{\mbox{mag/$\sq\arcsec$}}   

\received{2001 November 7}

\begin{document}

\slugcomment{Submitted for publication in {\it The Astrophysical
Journal}}

\title{Deep Near Infrared Mapping of Young and Old Stars 
in Blue Compact Dwarf Galaxies}

\author{Luz M. Cair\'os}
\email{luzma@uni-sw.gwdg.de}
\affil{Universit{\" a}ts-Sternwarte G{\" o}ttingen, 
Geismarlandstr. 11, 37083, G{\" o}ttingen, Germany. 
Instituto de Astrof\'\i sica de Canarias, V\'\i a L\'actea, E-38200  La Laguna, Tenerife,
Canary Islands, Spain. 
Departamento de Astronom\'\i a, Universidad de Chile, Casilla 36-D, Santiago de
Chile, 
Chile}

\author{Nicola Caon}
\email{ncaon@ll.iac.es}
\affil{Instituto de Astrof\'\i sica de Canarias, V\'\i a L\'actea, E-38200  La Laguna, Tenerife,
Canary Islands, Spain}

\author{Polychronis Papaderos}
\email{papade@uni-sw.gwdg.de}
\affil{Universit{\" a}ts-Sternwarte G{\" o}ttingen, 
Geismarlandstr. 11, 37083, G{\" o}ttingen, Germany}

\author{Kai Noeske}
\email{knoeske@uni-sw.gwdg.de}
\affil{Universit{\" a}ts-Sternwarte G{\" o}ttingen, 
Geismarlandstr. 11, 37083, G{\" o}ttingen, Germany}  

\author{Jos\' e M. V\'\i lchez}
\email{jvm@iaa.es}
\affil{Instituto de Astrof\' \i sica de Andaluc\'\i a, CSIC, Apdo. 3004,
18080 Granada, Spain}

\author{Bego\~na Garc\'\i a Lorenzo}
\email{bgarcia@ing.iac.es}
\affil{Isaac Newton Group of Telescopes, E-38780 Santa Cruz de La Palma,
La Palma, Canary Islands, Spain}

\author{Casiana Mu\~noz-Tu\~n\'on}
\email{cmt@ll.iac.es}
\affil{Instituto de Astrof\'\i sica de Canarias, V\'\i a L\'actea, E-38200  La Laguna, Tenerife,
Canary Islands, Spain}

\accepted{}
\shortauthors{Cair\'os et al.}
\shorttitle{NIR observations of BCDs}

\begin{abstract}  
We analyze $J$, $H$ and \ks\ near-infrared data for 9 Blue Compact Dwarf (BCD)
galaxies, selected from a larger sample that we have already studied in the
optical. We present contour maps, surface brightness and color profiles, as 
well as color maps of the sample galaxies.  
The morphology of the BCDs in the NIR has been found to be basically the same 
as in the optical. 
The inner regions of these systems are dominated by the starburst component. 
At low surface brightness levels the emission is due to the underlying host 
galaxy; the latter is characterized by red, radially constant colors and 
isophotes well fit by ellipses.
We derive accurate optical near--infrared host galaxy colors for eight 
of the sample galaxies; these colors are typical of an evolved stellar 
population.
Interestingly, optical near--infrared color maps reveal the presence of a
complex, large-scale absorption pattern in three of the sample galaxies. We
study the applicability of the S{\'e}rsic law to describe the surface 
brightness profiles of the underlying host galaxy, and find that, because of
the limited surface brightness interval over which the fit can be made, the
derived S{\'e}rsic parameters are very sensitive to the selected radial
interval and to errors in the sky subtraction. Fitting an exponential model
gives generally more stable results, and can provide a useful tool to quantify 
the structural properties of the host galaxy and compare them with those of 
other dwarf classes as well as with those of star-forming dwarfs at higher 
redshifts.
\end{abstract}


\section{Introduction}
\label{intro}

Blue Compact Dwarf (BCD) galaxies were first defined as compact and
low-luminosity objects (starburst diameter $\leq 1$ kpc; $M_{B} \geq -18$ mag),
with optical spectra similar to those presented by \ion{H}{2} regions in spiral
galaxies \citep{Sargent70,Thuan81}. Further work showed that they are
metal-deficient galaxies ---~the oxygen abundance of their ionized gas varying
between $Z_{\odot}/50$ and $Z_{\odot}/2$~--- which form stars at a rate that
would exhaust their gas content on a time scale much shorter than the age of
the Universe. Nowadays it seems clear that among the galaxies originally
classified as BCDs we find a wide range of luminosities ($-12 \geq M_{B} \geq
-21$ mag) and morphologies, the latter probably reflecting different
evolutionary states and star formation histories \citep{Loose86, Noeske00, 
Kunth00, Cairos01a, Cairos02,Bergvall02}. Initially it was hypothesized that
BCDs are truly young galaxies, forming their first generation of stars
\citep{Sargent70, Lequeux80, Kunth86, Kunth88}; the subsequent detection of an
extended, redder stellar host galaxy in the vast majority of them has revealed
that most BCDs are old systems \citep{Loose86, Telles95, Papaderos96a,
doublier97, doublier99, Cairos00, Cairos01a, Cairos01b, Bergvall02} undergoing 
recurrent star-forming episodes \citep{Thuan91,Mashesse99}.

BCDs are attracting a great deal of interest in astrophysical research, as it
has become clear that they could be the touchstones for understanding several
fundamental astrophysical problems. Even if most BCDs are not genuinely young
systems, their gas-richness and metal-deficiency make them useful objects for
constraining the primordial $^{4}$He abundance 
\citep{Pagel92,Masegosa94,Thuan95,Izotov97,Izotov99} and early metal enrichment
in galaxies. Furthermore, BCDs are ideal laboratories for the study of many
topics related to star formation (SF), by avoiding the complications of those
mechanisms, such as density waves, that operate in spiral galaxies
\citep{Oconnell78,Thuan91,Brosch98}. Last, it has been suggested that these
systems could be the nearby counterparts of the faint blue galaxies detected at
intermediate redshifts \citep{Koo94,Koo95,Guzman96,Guzman97}. Should this be
the case, BCDs offer the possibility of studying the properties of this galaxy
class with an accuracy and spatial resolution that cannot be achieved at larger
distances.

In spite of the interest that these systems have attracted, the  basic 
questions concerning their evolutionary status, formation history and
mechanisms that trigger their star-forming activity are still open. In
particular, the assessment of the properties of the stellar
low-surface-brightness component (LSBC) underlying the starburst regions  in
BCDs is a fundamental task. This older component hosts most of  the galaxy's
stellar mass.  Provided that Dark Matter (DM) does not dominate the mass
within  the Holmberg radius \citep{Papaderos96b}, the stellar LSBC is, together
with the \ion{H}{1} halo, mainly responsible for the global gravitational
potential within which the starburst phenomenon takes place. Moreover, deep
spectrophotometric studies of the LSBC are indispensable for putting
constraints on the evolutionary status and the formation  history of BCDs. The
comparison of the properties of the LSBC (e.g., structural parameters, average
colors and color gradients) with those of  other dwarf galaxy classes: dwarf
irregulars (dIs), dwarf ellipticals (dEs), and low surface brightness (LSB)
galaxies, will be crucial to test evolutionary scenarios that link those
galaxies with BCDs  \citep{Thuan85,davies88,Papaderos96b,Marlowe97,
Marlowe99,Cairos00}, or connect the different morphological subclasses of BCDs
\citep{Noeske00}. 

A precise determination of the structural properties of the LSBC is also
necessary for an elaborate study of the star-forming regions of BCDs
\citep{Telles95,Vanzi96,Telles97,Cairos00,Gildepaz00,Doublier01,Vanzi02,Papaderos02}.
Recently \cite{Cairos02} have compared the photometric parameters of the
individual star-forming knots in the BCD Mrk~370 before and after subtracting
the contribution from the older stars, and found significant differences 
in the results (the LSBC contribution typically shifts the starburst colors 
by 0.2--0.4 mag). Evidently such a correction must be computed and applied
before using the colors of the starburst knots to estimate their age.

The present study constitutes the third part of an extensive multiwavelength 
investigation of nearby BCDs. In the first stage of this program 
(Paper~I) optical broad-band 
images of 28 BCDs were used to examine the optical morphology and study the 
photometric structure of these systems by means of deep surface photometry.
In Paper~II we provided integrated photometry 
of the galaxies and produced an atlas of detailed color and \Ha\ maps. 

From the latter study it was concluded that the structural parameters  and the
evolutionary status of the stellar population underlying the  starburst
component can not be determined from optical information alone.  The optical
domain is strongly affected by interstellar reddening,  and dominated by the
emission of young stars and ionized gas all over  the inner part of a BCD.
Moreover, optical colors are not sensitive  age indicators for stellar
populations several Gyr old. For instance, for a single stellar population, the
\bv\ and \vr\ colors  increase by only $\approx 0.1$ mag when the age varies
from 5 to 15 Gyrs  (see, e.g., \citealp{Vazdekis96}).  The NIR spectral window
which traces primarily the old stellar populations  would be best suited for
studies focusing on the LSBC.  It offers the following advantages over the
optical:  
i) the contribution of young stars and nebular emission lines to the
total light of the galaxy is smaller than in the optical. The models by
\cite{kruger95} predict that a moderately strong burst at its peak in
luminosity accounts for only $\approx$ 20\% of a BCD's emission in the $K$-band,
compared to $\geq$ 80\% in $B$. Dealing with the BCD II~Zw~40, \cite{Vanzi96} found
that approximately one half of the total $B$ and $V$ emission of the
star-forming regions comes from emission lines, whereas the latter contribute
to less than 10\% in the $H$ and $K$ bands. There is also observational
evidence that the starburst fades away at galactocentric distances smaller 
than in the optical \citep{alton94,james94,Vanzi96,beck97}.
ii) NIR broad-band colors are much less affected by interstellar  extinction,
the extinction in $H$ being a factor of $\sim 6$ lower than in $V$.  
iii)  A combination of optical-NIR colors allow, in principle, to distinguish
among more or less evolved populations of old stars 
\citep{Thuan85,Oestlin98,Peletier90,Peletier99}.

Consequently, the next stage in our project has been to extend our analysis to
the NIR regime. Here we report on deep surface brightness photometry in 
$J$, $H$ and $K_{s}$ for a sample of 9 BCD galaxies, 8 of which have 
been previously analyzed by us in the optical (see Papers~I and ~II).

\section{Observations, Data Reduction and Flux Calibration}
\subsection{Observations}
\label{obser}

The basic data for the sample galaxies are given in Table~\ref{Tab:log}. Column
5 lists the  galaxies distances, which were computed assuming a Hubble flow, 
with a Hubble constant $H_{0}$ = 75 km s$^{-1}$ Mpc$^{-1}$, after correcting
recession velocities relative to the centroid of the local group for 
virgocentric infall. Column 6 lists the absorption coefficient in the $B$
band.  Absolute magnitudes (column 7) have been obtained from the $B$
asymptotic  magnitudes \citep{Cairos00}, using the distances tabulated in
column 5.  The $B$ magnitude of Mrk~407 has been obtained from NED.

NIR imaging of the galaxies in $J$, $H$ and \ks\ was obtained in April 2000 at 
the 4.2m  William Herschel Telescope (WHT, Observatorio del Roque de los 
Muchachos, ORM, La Palma) during three consecutive nights. We used the 
infrared camera INGRID (Isaac Newton Group Red Imaging Device), equipped with 
a 1024$\times$1024 pixel Hawaii detector array.  The scale was 0\farcs 242
pixel$^{-1}$ and the field of view was  $4\farcm 13 \times 4\farcm 13$.  We
preferred to use the \ks\ instead of the standard $K$ filter, as the former 
has a more favorable transmission curve, which cuts off more efficiently the 
thermal emission from the telescope and the atmosphere, thus reducing 
considerably the background noise in the final images. Throughout this work we
adopt $K=\ks$, which turns out to be a reasonable assumption
\citep{Persson98,Gonzalezperez01}.

During the first two nights of the run the seeing was quite stable, varying
between 0\farcs 8 and 1\farcs 2 FWHM, while during the third night it 
deteriorated to values between 1\arcsec\ and 1\farcs 8.

The galaxies Mrk~36, II~Zw~70, II~Zw~71 and Mrk~297 were also observed 
in April 1999 at the Centro Astron{\'o}mico Hispano Alem{\'a}n (CAHA) at Calar 
Alto with the 1024$\times$1024 pixel OMEGA PRIME camera, mounted at the 3.5m 
telescope. 
The scale was 0\farcs 3961 pixel$^{-1}$, and the total field of view $6\farcm
76\times 6\farcm 76$. The targets were imaged in the $J$, $H$ and $K_{m}$
bands ($K_{m}$ is, like \ks, a modified $K$ filter that reduces the thermal 
background; \citealp{Wainscoat92}). 
The seeing varied between 1\farcs 2 and 2\farcs 2 FWHM.
A complete log of the observations is provided in Table~\ref{Tab:logobs}.

\begin{deluxetable}{lcccccc}
\tabletypesize{\footnotesize}
\tablewidth{0pt}
\tablecaption{Sample of Galaxies\label{Tab:log}}
\tablehead{\colhead{Galaxy} & \colhead{Other designations} &
\multicolumn{2}{c}{R.A.\ \ (2000) \ \ Dec.} &
\colhead{D (Mpc)} & \colhead{A$_{B}$ (mag)} & \colhead{M$_{B}$ (mag)} }
\startdata
Mrk~407         &                      & 09 47 47.6  & \phs39 05 04  &  23.2  & 0.069  & $-16.33$  \\[3pt]
Mrk~33          & Haro~2, Arp~233      & 10 32 31.9  & \phs54 24 03  &  22.3  & 0.052  & $-18.28$  \\[3pt]
Mrk~35          & Haro~3, NGC~3353     & 10 45 22.4  & \phs55 57 37  &  15.6  & 0.031  & $-17.75$  \\[3pt]
Mrk~36          & Haro~4               & 11 04 58.5  & \phs29 08 22  &  10.3  & 0.131  & $-14.69$  \\[3pt]
UM~462          & Mrk~1307             & 11 52 37.3  & $ -02$ 28 10  &  14.0  & 0.083  & $-16.07$  \\[3pt]
II~Zw~70        & Mrk~829              & 14 50 56.5  & \phs35 34 18  &  19.0  & 0.053  & $-16.55$  \\[3pt]
II~Zw~71        &                      & 14 51 14.4  & \phs35 32 31  &  19.6  & 0.055  & $-17.00$   \\[3pt]
I~Zw~123        & Mrk~487              & 15 37 04.2  & \phs55 15 48  &  12.5  & 0.062  & $-15.04$   \\[3pt]
Mrk~297         & NGC~6052, Arp~209    & 16 05 12.9  & \phs20 32 32  &  65.1  & 0.330  & $-21.01$   \\
\enddata
\end{deluxetable}

\begin{deluxetable}{lccccc}
\tabletypesize{\footnotesize}
\tablewidth{0pt}
\tablecaption{Log of the observations\label{Tab:logobs}}
\tablehead{\colhead{Galaxy} & \colhead{Date} & \colhead{Telescope} &
\multicolumn{3}{c}{Exposure time (s)}\\[1pt]
\colhead{\mbox{}}&\colhead{\mbox{}} & \colhead{\mbox{}} & \colhead{$J$} & \colhead{$H$} & \colhead{\ks}
}
\startdata
Mrk~407         & Apr.  00  & WHT 4.2m     &  1200   &  1600    &  2205     \\[3pt]
Mrk~33          & Apr.  00  & WHT 4.2m     &  1700   &  1200    &  2565     \\[3pt]
Mrk~35          & Apr.  00  & WHT 4.2m     & \phn800 &  \phn720 &  1440     \\[3pt]
Mrk~36          & Apr.  00  & WHT 4.2m     &  1200   &  \phn240 &  \phn120  \\
                & Apr.  99  & CAHA 3.5m    & \nodata &  1500    &  2250     \\[3pt]
UM~462          & Apr.  00  & WHT 4.2m     &  1200   &  1280    &  1800     \\[3pt]
II~Zw~70        & Apr.  00  & WHT 4.2m     &  1700   &  \phn640 &  \phn540  \\
                & Apr.  99  & CAHA 3.5m    &  2280   &  \phn420 &  \phn990  \\[3pt]
II~Zw~71        & Apr.  00  & WHT 4.2m     & \phn960 &  \phn720 &  \phn480  \\
                & Apr.  99  & CAHA 3.5m    &  1080   &  \phn360 &  1320     \\[3pt]
I~Zw~123        & Apr.  00  & WHT 4.2m     &  1300   &  1600    &  1440     \\[3pt]
Mrk~297         & Apr.  00  & WHT 4.2m     & \phn600 & \phn420  &  \phn360  \\
                &          & CAHA 3.5m     & \nodata &  1200    &  \phn600  \\[3pt]
\enddata
\end{deluxetable}

\subsection{Data Reduction}
\label{datare}

The high sky background levels in the NIR (especially in $H$, because of OH
airglow and in $K$, because of thermal emission from sky, telescope and dome)
make the observing strategy and the data reduction process more complicated
than when observing in the optical. The telescope must chop between on-target
and off-target positions to properly sample variations of the sky-background.
Additionally, the total integration time spent at each of these positions must
be split into many sub-exposures of a few seconds each in order not to
exceed the linearity range of the detector. 

The individual exposure times in INGRID were typically 50 s in $J$, 40 s in
$H$ and 15 s in \ks, with sets of 2--4 exposures taken at each position.
The total on-target integration times are listed in Table~\ref{Tab:logobs}.
For the CAHA observations, the telescope position was changed between pointings
of 60 s each, which were split up into sub-exposures of 6, 3 and 2 s in
$J$, $H$ and $K_m$, respectively. 

The processing and calibration of the frames was performed using standard
procedures available in IRAF\footnote{IRAF (Image Reduction and Analysis
Facility) is distributed by the National Optical Astronomy Observatories, 
operated by the Association of Universities for Research in Astronomy, Inc.,
under cooperative agreement with the National Science Foundation.}. 

We first built a bad pixel mask, which was applied to every raw frame.  This
mask is used to correct for bad pixels by interpolating the flux  level along
neighboring lines and columns. Second, the sky background has to be subtracted
from each frame. A typical NIR frame contains flux contributions from the
target, from the sky background, from the warm telescope and from dome
scattered light. Hence, it is necessary to take sky frames to measure and
remove these contributions, which may fluctuate on time-scales of minutes.
Depending on the size of the target, we have employed two different  procedures
in order to sample the background: 1) when dealing with small objects (optical
diameter $\leq 30$\arcsec), we offset the telescope between exposures,
producing a sequence in which the target is placed on different sections
(typically the four quadrants) of the array. The median average of these frames
will then provide a clean sky frame (free from the  target, cosmic rays and
point-like sources). This procedure has the benefit that all observing time is
spent on the target. 2) The above technique cannot be used when observing large
objects which  extend over a significant portion of the detector. In those
cases, in order to  obtain a clear patch of sky, we offset the telescope some
arcmin from the  target position, each time dithering by several arcsecs.

Afterwards, the sky subtracted frames must be flat-fielded. To create a proper
flat-field, which contains only information on the overall response and
pixel-to-pixel sensitivity variations of the telescope/camera combination,  two
type of flat-field images are needed: the "flat on", with a high count  level,
and the "flat off", with the same duration of the "on" but with a low  count
level. In the INGRID run, flats were taken during twilight  (``flat on''
when the sky was still bright, ``flat off'' when the sky was  darker). In the
CAHA run, the flat-fields were obtained by taking exposures of the dome  with
the lamps on and off. The subtraction of the dark flat from the bright one
(after averaging out the several individual frames) removes the dome, telescope
and dark-current  contributions, giving the final flat-field image. 

The sky-subtracted target frame is then divided by the flat-field and,
finally, all frames in the same band are aligned, brought to the same
resolution by convolution with a gaussian of appropriate width, and combined.

In the case of II~Zw~70 and II~Zw~71, we built the final frames by combination 
of the images taken at CAHA and at the WHT. The WHT images were resampled 
and convolved by a gaussian to match the same pixel size and actual resolution
of the CAHA frames; finally they were averaged weighting them by their
signal-to-noise ratio.

\subsection{Flux Calibration}
\label{fluxca}

Flux calibration was done through the observation of photometric standard stars
selected from the UKIRT list of faint IR standard stars. The stars FS~17,
FS~21, FS~27 and FS~35 were observed throughout each night at different
airmasses. The range of colors in the NIR is very small for normal stars
(particularly \hk) and, as a consequence, accurate determination of the color
term could not be achieved using the standard list. Thus, when calculating the
calibration constant and the magnitudes of the galaxies, we neglect the color
term and resolve the following equation:

\begin{equation}
m_{\lambda_{0}}=m_{\lambda_{i}}+ a_{0} + a_{1}X,
\end{equation}

\noindent
where $m_{\lambda_{0}}$ is the magnitude in the standard system,
$m_{\lambda_{i}}$ is the instrumental magnitude, $X$ is the airmass at which
the observations were done and $a_{i}$ are the calibration constants to be
determined.

The observations taken at CAHA were calibrated using the INGRID frames by
matching the fluxes computed within a sequence of centered apertures. The
magnitudes in $K_s$ and $K_m$ were transformed following \cite{Wainscoat92}. 
Note that the $H$ and $K_s$ INGRID images of Mrk~36 and Mrk~297 were used only 
to calibrate the CAHA exposures.

\section{Photometry and Morphology of the Galaxies} 
\subsection{Isophotal magnitudes}
\label{isomag}

In order to allow a direct comparison of our photometry with values derived in 
other studies, we list in Table~\ref{Tab:IntPhot} the isophotal magnitudes of 
the sample BCDs in $J$, $H$ and $K_s$, as respectively determined within the 
23, 22 and 21 mag arcsec$^{-2}$ isophote.
These magnitudes are tabulated alongside with the uncertainties resulting from 
the zero-point calibration and the subtraction of the sky background.

Integrated magnitudes were corrected for Galactic extinction by using the
absorption coefficient in the $B$ band ($A_{B}$) and the color excess $E(\bv)$
derived from \citet{schlegel98}. The absorption coefficients in the NIR bands
($A_{J}$, $A_{H}$ and $A_{K}$) are derived by assuming the standard extinction
law ($R_{V} = A_{V}/E(B-V) = 3.1$), and using the $A_{\lambda}$ curve from
\citet{Cardelli89}. The values of $A_{J}$ for each object are also listed in 
Table~\ref{Tab:IntPhot}.

\begin{deluxetable}{lcccc}
\tabletypesize{\footnotesize}
\tablewidth{0pt}
\tablecaption{Isophotal magnitudes \label{Tab:IntPhot}}
\tablehead{\colhead{Galaxy} & \colhead{A$_{J}$ (mag)} & 
           \colhead{$J$ (mag)} & \colhead{$H$ (mag)} & \colhead{\ks\ (mag)} 
           } 
\startdata
Mrk~407      & 0.014 &  $13.28\pm0.04$ & $13.03\pm0.04$ &  $12.66\pm0.05$  \\
Mrk~33       & 0.011 &  $11.28\pm0.02$ & $10.84\pm0.03$ &  $10.65\pm0.04$  \\
Mrk~35       & 0.006 &  $11.31\pm0.03$ & $10.83\pm0.02$ &  $10.43\pm0.03$  \\
Mrk~36       & 0.027 &  $14.29\pm0.03$ & $14.03\pm0.03$ &  $13.95\pm0.04$  \\
UM~462       & 0.017 &  $13.32\pm0.03$ & $12.87\pm0.02$ &  $12.72\pm0.03$  \\
II~Zw~70     & 0.011 &  $13.42\pm0.05$ & $13.05\pm0.04$ &  $13.00\pm0.05$  \\
II~Zw~71     & 0.012 &  $12.62\pm0.06$ & $12.14\pm0.03$ &  $12.16\pm0.05$  \\
I~Zw~123     & 0.012 &  $13.76\pm0.02$ & $13.25\pm0.03$ &  $13.20\pm0.04$  \\
Mrk~297      & 0.069 &  $11.37\pm0.02$ & $10.92\pm0.06$ &  $10.58\pm0.06$  \\
\enddata
\tablecomments{Column 2: Galactic extinction in $J$;
(3)--(5): isophotal magnitudes in the different bands (within 23 \sbb\ in $J$, 
22 \sbb\ in $H$ and 21 \sbb\ in $K_s$), corrected for galactic extinction.}
\end{deluxetable}

\subsection{Contour and Color Maps}
\label{comap}

Figure~\ref{Fig:SBP1} (top panel, on the left) show the $J$
band contour maps of the sample galaxies ($H$ band for Mrk~297). In order to
better trace the lower surface brightness levels, the outer contours were
computed after smoothing the original image by a boxcar sliding average, with
a window size increasing with the distance to the center (typically up to
$7\times7$ or  $9\times9$ pixels).

We find that the morphology of the galaxies is usually very similar in the
optical and in the NIR (see the atlas of $B$ band contour maps in Paper~I). 
Because of the better spatial resolution in the NIR, in several cases (UM~462
being the most striking) we can resolve knots of SF which appeared blurred in
the optical frames.  Also, because our NIR data do not reach surface
brightness levels as deep as in the optical, some galaxies (for instance,
Mrk~36 and UM~462), appear slightly irregular in their outer part.

Using the NIR images, we grouped the galaxies into the four different
morphological classes defined in Paper~II and confirm in all cases the 
classification previously made on optical frames.

We have also computed optical-NIR color maps of our sample BCDs. 
These maps were built after the calibrated frames have been aligned,
brought to the same pixel scale (the one in the frames 
with the poorer resolution) and, finally, seeing-adapted. 
Usually the optical-NIR color maps are limited by the noise in the 
NIR frames, which is considerably higher than in the optical, so reliable 
information can be only derived for the central, high intensity galaxy regions.

The \vk\ color maps of the galaxies (\bk\ in the case of II~Zw~71) are
displayed for the galaxies with the most interesting morphology in 
Figures~\ref{Fig:VKmkn33colormaps} to \ref{Fig:VKmkn297colormaps}. 
Extended dust patches are clearly visible in three galaxies: Mrk~33, Mrk~35 
and Mrk~297.

\subsection{Surface brightness, color and shape profiles}
\label{surpho}

Figure~\ref{Fig:SBP1} display the surface brightness
profiles (SBPs) of the galaxies in the three NIR bands. We also present
the numerical ellipticity and position angle distribution in the $J$ band
as a function of the equivalent radius $R$, as well as color profiles (CPs) 
in the NIR. Whenever optical information is available, we also computed
optical-NIR CPs. 

SBPs were built following the methods described in \citet{Cairos00} and in
Paper~I, which we summarize here. In the inner, irregular starburst regions,
where the standard technique of fitting ellipses to the isophotes can not be
applied,  we used an alternative method, which does not require any assumption
on the  morphology of the object. We define $A_{n}$ as the sum of the areas of
all those pixels  the intensity of which is higher than $I_{n}$:

\begin{equation}
A_{n}=N_{{\rm pix}, n} A_{\rm pix},
\end{equation}

\noindent
$N_{{\rm pix}, n}$ is the number of the pixels with $I \geq
I_{n}$ and  $A_{\rm pix}$ is the pixel area in square arcseconds.

The flux $F_0$, $F_1$, {\ldots}, $F_n$ enclosed by the isophotes with 
intensities  $I_{0}$, $I_{1}$, ...., $I_{n}$ (spaced 0.1 mag apart), enclosing 
areas  $A_{0}$, $A_{1}$, ..., $A_{n}$, is then calculated by performing 
consecutive integrations over the image. To each annulus we associate 
a radius $R_{n}$, equal to the radius of the circle whose area is the 
average of $A_{n}$ and $A_{n+1}$:

\begin{equation}
R_{n}=((A_{n}+A_{n+1})/2\pi)^{1/2},
\end{equation}

\noindent
while the surface brightness $\mu_n$ (\sbb) is obtained by dividing the flux 
within each annulus by its area:

\begin{equation}
\mu_n = -2.5\;\log [(F_{n+1}-F_{n})/(A_{n+1}-A_{n})] + {\rm const},
\end{equation}

\noindent
where const is the calibration constant.
The surface brightness is then plotted versus the equivalent radius.
Outside the starburst-dominated regions, our sample BCDs present a relatively
regular and symmetric morphology. 
We derived the luminosity and geometrical profiles by fitting ellipses to the 
isophotes, using the {\sc iraf} task {\em ellipse}.
First we masked out foreground stars, smaller knots, and other disturbances. 
Then we ran the task on the original image and on images obtained by smoothing
the former by a median average (with window sizes up to $11\times 11$ pixels),
to improve the stability and reliability of the results in the outer galaxian
regions.

The final surface brightness and geometrical profiles were then built by 
combining the profiles derived from the un-smoothed image in the inner parts 
with those derived from the smoothed images in the outer parts, checking that 
the profiles match well in the overlapping radial ranges.

The errors of the surface brightness levels are computed as the quadratic sum,
in intensity, of the uncertainties given by the ellipse fitting algorithm and
those in the sky subtraction. The latter (which is not to be confused with the
noise of the sky), is an estimate of how much the sky level varies around the
galaxy as a result of large- or intermediate-scale residual gradients in the
sky background, imperfect flat-fielding, etc.

Color profiles have been derived by subtracting the SBPs in the two bands,
after the SBPs have been brought to the same resolution using gaussian
smoothing and resampled to equal radial step. We note that in their very center
CPs can still present sudden variations, as the SBP extraction technique
employed at high-surface brightness levels where multiple star-forming knots
are present does not allow to perfectly reproduce the  color variations from
knot to knot. In such areas two irregular "annuli" having the same equivalent
radius in the two bands may have different shapes, so we might compare the
signal from physically different spatial regions. Thus color profiles are only
plotted outside the central 1--2\arcsec. 

We have found that, in agreement with the optical results, one can usually
distinguish on near-infrared SBPs two components: the starburst and the LSBC
host galaxy. This was found to be the case for 8 out of the 9 sample BCDs. Only
in Mrk~407 we cannot separate the two components. However, given the faintness
and compactness of the galaxy, we can neither prove nor rule out the presence
of a LSBC;  profiles reaching deeper surface brightness levels are clearly
necessary to give a firm answer.

In all cases the SBPs have the same shape as the optical profiles (see Figure~1
in Paper~I); however, the contribution of the starburst in the NIR is much less
important than in the optical: the intensity peaks are less pronounced and the
emission appears to be less extended (see also \citealp{Noeske03}). For
UM~462 the NIR profiles are, in their inner part, markedly steeper that the 
optical SBPs presented in Paper~I, because of the much better resolution of 
the NIR data.

\section{Structure and colors of the underlying host}
\label{hoststru}

Surface photometry studies in the optical have clearly shown that the SBP of
BCDs is the superposition of at least two components: the starburst, made up of
the emission from young massive stars and from ionized gas, which dominates the
light at high and intermediate intensity levels, and the stellar low surface
brightness component, which dominates at larger radii and contains most of the
system's stellar mass 
\citep{Loose86,Telles95,Papaderos96a,Marlowe97,doublier97,Oestlin98,
doublier99,Cairos01a,Cairos01b,
Bergvall02}.

However, not much in-depth work has been done so far on the properties of the
LSBC, mainly because of two reasons: first, only recently it has been
recognized that its structural properties may significantly influence the
global star formation process in BCDs \citep{Papaderos96b,Cairos02} and that
its morphology may hold clues to their evolutionary state
\citep{Oestlinber98,doublier97,doublier99,Noeske00,Kunth00,Cairos02};
second, because of its faintness, a quantitative study of the LSBC (derivation
of, e.g., scale lengths, total luminosities or color gradients) requires a
great deal of observational and analysis effort.

The purpose of this section is to derive the NIR structural parameters of the 
LSBC for the observed galaxies. The proper determination of these parameters 
requires some caution. In fact, in BCDs the emission from the young stars and
the ionized gas camouflages the LSBC at high and intermediate intensity levels;
it is, therefore, fundamental to identify and measure the size of the region
contaminated by the starburst, and derive the structural parameters of the LSBC
\textit{outside} this region . Since this means to perform the fit usually for
$ \mu_B \geq 24$ mag arcsec$^{-2}$  \citep[][
Paper~I]{Papaderos96a,Bergvall02}, it is essential to achieve deep surface
brightness limits: the deeper the data, the larger the portion of the host
galaxy that we can analyze, and therefore the more reliable the final results. 
In particular, three are the main requirements:

a) {\em Proper determination of the fit radial range}. Selecting the radial
range of the SBP in which the fit is to be done is possibly the most critical
point: we must make sure that we fit uniquely those regions of the galaxy
free from starburst emission. In order to delimit this region, we took
advantage of our optical data. By inspecting the optical color maps and \Ha\
images of the galaxies (Paper~II), we defined an $R_{\rm transition}$ outside
which the starburst emission is practically absent. The fit was then done in
the interval $R > R_{\rm transition}$. We notice that an examination of the SBP
itself is not sufficient: in fact, the SBP might show no discontinuity or
marked change of slope around $R_{\rm transition}$, and one may be tempted to
extend inwards the fit interval, which makes the derived LSB structural
parameters significantly skewed by the inclusion of starburst light.

b) {\em Deep and extended surface brightness
 profiles}.  To fulfill it, we observed at 4m
class telescopes with long exposures times, and did a careful flat-field
correction and sky background subtraction, both crucial steps in order to reach
the required surface brightness levels (27-28 \sbb in $B$, 24-25 \sbb in $J$).

c) {\em Use of a suitable analytical model for the light distribution of the
host}. A third important point is the choice of the most appropriate formula to
describe the SBP of the host. So far, it has been common practice to adopt the
exponential model. However, it is nowadays recognized that several LSB host
have profiles that show systematic deviations from the exponential model. Here,
we go further and explore the applicability of the S{\'e}rsic law in
parametrizing the radial intensity  distribution of the underlying LSBC of
BCDs.

Underlooking or underestimating the importance of the above points can lead to
incorrect determinations of the LSB structural parameters, and may explain the
large discrepancies among the results obtained in the optical bands by
different authors (see the discussion in\citealp{Kunth00,
Cairos00,Cairos01a}).

\subsection{Exponential and S{\'e}rsic models}
\label{models}

So far, it has been common practice to model the host galaxy of BCDs by 
fitting an exponential function to the outer parts of SBPs 
\citep{Telles95,Papaderos96a,Marlowe97,Oestlin98,Cairos01a}.

\begin{equation}
I(R)=I_{0}\exp(-R/\alpha), 
\label{expo}
\end{equation}

\noindent
where $I_{0}$ is the central intensity and $\alpha$ is the scale length. 
Equation \ref{expo} corresponds to a straight line in the $R-\mu$ diagram. Its
total magnitude, $m_{\rm T}$, is then given by: 
$m_{\rm T}=\mu_0-5\;\log(\alpha)-1.995$. 
However, it is clear that several SBPs show a curvature in the $R-\mu$ diagram 
and, thus, the derived exponential fit parameters may depend on the radial 
range where the fit is done (see, for instance, the SBP of Mrk~370 and Mrk~33, 
Figure~2, Paper~I). 
\citet{Cairos00} has observed a concave outer slope in several deep optical 
SBPs, and remarked that in those cases, a seemingly irreproachable exponential 
fit to the LSBC emission could just be due to fitting a small portion of the 
SBP. 
Conversely, in some BCDs an inwards extrapolation of the  exponential slope of 
the LSBC implies for intermediate to small radii an intensity higher than the 
observed value. This fact, which is suggestive of an overall convex LSBC 
profile, has been  discussed in, e.g., \citet{Papaderos96a}; these authors 
(see also \citealt{Noeske03}) proposed a modified exponential distribution 
flattening in its inner part to fit SBPs presenting the above mentioned 
characteristics.

In the last years, the S{\'e}rsic law has grown in popularity as the most
appropriate description of the SBP of ellipticals and bulges:

\begin{equation}
I(r)=I_{e}\exp(-b_n[(r/r_{e})^{1/n}-1]), 
\label{ser}
\end{equation}

The coefficient $b_n$ is chosen so that $r_e$ is the radius enclosing half of
the total light of the model and $I_{e}$ denotes the intensity at $r_e$. The
above formula includes as particular cases the exponential  ($n=1$) and the de
Vaucouleurs ($n=4$) laws. \citet{Caon93} showed that the $n$ parameter is
directly correlated to the size/luminosity of elliptical galaxies. This finding
has been extended to brightest cluster members \citep{Graham96}, bulges of
spirals \citep{Andredakis95} and dwarf ellipticals \citep{YoungCurrie94}. A
physical interpretation of the shape parameter $n$ has been proposed by
\citet{TGC2001}, in the sense that a larger $n$ corresponds to a higher
central light (and possibly mass) concentration. 
Notably, the light concentration also correlates with the central stellar 
velocity dispersion $\sigma$ and  with the mass of the central super-massive 
black-hole \citep{GECT2001}.

It is then natural to ask whether the LSBC in BCDs can be described by a
S{\'e}rsic law too, and whether the index $n$ correlates with other properties
of the LSBC, such as total luminosity, scale length or central surface
brightness. Here we study the shape of NIR profiles in their outer regions, by
fitting a S{\'e}rsic law. We also investigate how the resulting model
parameters may be influenced by the precise radial range in which the profile
is fitted and by errors in the sky subtraction.

\subsection{On fitting a S{\'e}rsic law to the SBP of the underlying 
host galaxy}
\label{Fitsersic}

As we shall demonstrate below, fitting S{\'e}rsic models to the LSBC emission
in BCDs is not as straightforward as one may think, and the result may be
entirely dictated by observational uncertainties. For the determination of $n$
for a real SBP to be reliable, the  fitting range must be quite extended,
spanning a surface brightness interval of at least 4 magnitudes. This is not a
problem when monocomponent galaxies are fitted, or when careful bulge-disk
decomposition is done in deep images\footnote{For instance, in \citet{Caon93}
the typical range in surface brightness was 5--6 mag.}. 
However, these requirements can hardly be met in BCD galaxies, where, as
remarked earlier, the emission from the old stellar population in the central
regions is overwhelmed by the starburst. For instance, in the $B$ band the
starburst emission affects the SBPs down to typically  $\mu_B \ga 24$ mag
arcsec$^{-2}$  \citep[][ Paper~I]{Papaderos96a}, thus leaving only a range of
at most 3 or 4 magnitudes on which to fit a S{\'e}rsic model to the outer part
of a SBP. At these faint surface brightness levels, minor uncertainties in the
sky subtraction or image flat-fielding can distort the SBP shape and strongly
bias the derived parameters of Eq. \ref{ser}.

Also, the S{\'e}rsic parameter $n$ is poorly constrained when the fit is done
on such limited surface brightness intervals \citep[see][]{Makino90}. Small
variations in the fitted radial range, and the inclusion or exclusion of a few
datapoints may change the resulting $n$ more than one would naively expect. 
In principle, the situation in the NIR domain should be more favorable, 
because the starburst contaminates the light to a lesser degree. 
However, because of the high sky background, the S/N ratio in the NIR data 
is lower than in the optical (especially in $K$), so it is very difficult to 
obtain SBPs as deep and extended as in the optical bands. 

Because in several cases the BCD's profiles show a relatively smooth curvature
over a large radius range, one is tempted to enlarge the fit interval by 
decreasing $R_{\rm transition}$, at the risk of including regions still
affected by starburst light. In those cases the rms (the average scatter of the
residuals between observations and fit) is merely a poor indicator for the
validity of the fit solution, as it may marginally vary despite huge variations
in the derived $n$. The steeply increasing flux contribution of the starburst
towards small radii can indeed mimic a S{\'e}rsic profile with a higher $n$
($>4$, see, e.g. the discussion in \citealt{Papaderos96a}).

\subsubsection{Illustrative examples: Mrk~33 and Mrk~35}
\label{examples}

In order to illustrate the above problems, and provide some quantitative
estimates of how they could affect the derived S{\'e}rsic parameters, we carry
out a series of fits to the LSBC light profile of the BCDs Mrk~33 and Mrk~35.
For both we have deep, good quality data. Mrk~33 exhibits a smooth intensity
decay over its whole radius range, so that one may be tempted to fit its SBP
using a single S\'ersic law. The $R_{\rm transition}$ of this system can only
be inferred from color or H$\alpha$ maps, i.e. cannot be read off its SBP. On
the other hand, in Mrk~35 $R_{\rm transition}$ is approximately equal to the
radius where a sharp change in the SBP slope occurs, outside which the LSBC
shows an exponential behavior.

A typical approach to assess the uncertainties on the parameters obtained from
fitting models to galaxy light profiles is to create first a 2-D galaxy images
based on the selected analytical formula. From it, a number of simulations are
built by adding noise, convolving by seeing, over or undersubtracting the sky
background, etc., and the light profile is then extracted and fitted; the
difference between measured and input parameters is recorded. From the
distribution of these differences the parameter uncertainty is then estimated.
By repeating the above procedure for many combinations of input parameters,
their uncertainties can be studied as function of, for instance, the total
galaxy magnitude, or effective radius, or profile shape, etc.

However in the reality there are further and somewhat more complex source of
uncertainties often neglected in the above scheme, so that these uncertainties
may be actually under-estimated. For instance, the galaxy profile may be
contaminated by small knots or other disturbances not adequately masked out
before the ellipse fitting algorithm is run, or may be distorted by the
presence of nearby bright stars. Also, the sky background may be inhomogeneous,
presenting fluctuations on scales of the order of the galaxy size.\footnote{For
our data, seeing is not that important,  as $R_{\rm transition}$ is generally
larger than the region affected by seeing convolution.}

So in the present paper we use our observed data to carry out a more realistic
analysis of parameter uncertainties, based on comparing the parameters
obtained from images taken in different bands and by fitting in different 
radial intervals.

\subsubsubsection{Dependence on the radial range}

\label{Sect:mrk33sersic}

We fitted a S{\'e}rsic law to the SBP of Mrk~33 and Mrk~35 in different radial 
intervals; in these and all subsequent fits, we assigned the same weight
to all datapoints. This is to avoid that only a few points in the higher
surface brightness portion of the profile dominate the fit, which could 
be almost insensitive to the outermost profile datapoints.

The S{\'e}rsic parameters thus obtained are summarized in 
Table~\ref{Tab:fitmrk335}.

\begin{deluxetable}{llcccccccc}
\tabletypesize{\footnotesize}
\tablewidth{0pt}
\tablecaption{S{\'e}rsic parameters for Mrk~33 and Mrk~35 as function of the 
radial range\label{Tab:fitmrk335}}
\tablehead{\colhead{Fit ID} & \colhead{Galaxy} & \colhead{Band} & 
           \colhead{Radial range} & 
           \colhead{Surface brightness range} & \colhead{$n$} &
           \colhead{rms} & \colhead{$r_e$} & \colhead{$\mu_e$} &
           \colhead{$m_{\rm LSBC}$}  \\
	   \colhead{}        & \colhead{}     & \colhead{} &
	   \colhead{arcsec}       &
	   \colhead{\sbb}                     & \colhead{} &
	   \colhead{mag}   &  \colhead{arcsec}  & \colhead{\sbb}  &
	   \colhead{mag}
	   } 
\startdata
(1)  & Mrk~33 &$J$&  20.47--49.90 & 21.23--23.68 & $2.09$  & $0.041$  & 17.76 & 20.99 & 11.68\\
(2)  &        &   &  10.25--49.90 & 19.71--23.68 & $4.07$  & $0.038$  & 11.87 & 20.03 & 11.26\\
(3)  &        &   &  21.29--39.12 & 21.40--22.91 & $3.73$  & $0.030$  & 15.41 & 20.67 & 11.38\\
(4)  &        &   &  20.47--39.12 & 21.23--22.91 & $6.98$  & $0.031$  & 09.80 & 19.62 & 10.98\\
(5)  &        &   &  10.25--39.12 & 19.71--22.91 & $7.98$  & $0.024$  & 08.98 & 19.42 & 10.90\\
(6)  &        &\ks&  20.21--34.84 & 20.67--22.28 & $1.29$  & $0.048$  & 15.78 & 20.14 & 11.35\\
(7)  &        &   &  10.48--34.84 & 18.98--22.28 & $5.98$  & $0.052$  & 05.26 & 17.48 & 10.27\\[4pt]
(8)  & Mrk~35 &$J$&  11.54--38.64 & 19.71--23.58 & $0.87$  & $0.081$  & 13.70 & 20.05 & 11.73\\
(9)  &        &   &  15.45--38.64 & 20.32--23.58 & $0.76$  & $0.088$  & 14.61 & 20.23 & 11.84\\
(10) &        &   &  21.25--34.53 & 21.12--22.82 & $0.95$  & $0.105$  & 12.91 & 19.84 & 11.62\\
(11) &        &\ks&  12.67--43.70 & 19.15--23.12 & $0.84$  & $0.069$  & 15.10 & 19.43 & 10.92\\
(12) &        &   &  15.18--43.70 & 19.37--23.12 & $0.86$  & $0.075$  & 14.98 & 19.40 & 10.90\\
(13) &        &   &  21.95--34.54 & 20.19--22.05 & $0.73$  & $0.089$  & 15.32 & 19.44 & 10.96\\
\enddata 
\tablecomments{Columns 4, 5: radial interval and corresponding surface brightness
interval for the S{\'e}rsic fit; 
col 6: best-fit shape parameter $n$; 
col 7: rms scatter of the fit;
cols 8, 9: best-fit effective radius and effective surface brightness; 
col 10: total magnitude of the LSBC derived from the fit.} 
\end{deluxetable}

\begin{itemize}

\item {\bf Mrk~33} --- Optical studies (Paper I) show that starburst emission 
is limited to the central $16\arcsec$, so the $J$ band SBP can be safely
fitted in the range $20\arcsec <R<50\arcsec$, 50\arcsec\ being the last
observed datapoint; the fitting range corresponds to a surface brightness
interval of $\Delta \mu = 2.4$ mag.  Fitting Eq. \ref{ser} gives $n=2.09$ with
${\rm rms} = 0.041$. The smoothness and nearly constant curvature of the 
profile tempts to enlarge the fitted interval by extending it inwards to
$R=10$\arcsec\ ($\Delta \mu \sim 4$ mag); this gives $n=4.07$ and an improved
${\rm rms}=0.038$. As evident from comparison of rows 3 and 4 in 
Table~\ref{Tab:fitmrk335}, the exclusion of one single datapoint from the fit
can change $n$ by nearly a factor of two.

A similar behavior is found when fitting  Eq. \ref{ser} to the \ks\ band 
profile using the same starting radius (the \ks\ profile is less extended). 
In this case we obtain $n=1.29$ ($\Delta  \mu$ = 1.6 mag, ${\rm rms} = 0.048$),
fitting from $R = 20\farcs 20$, and $n=5.98$ fitting from $R = 10\farcs 48$, 
with only a very small increase in the rms.

In Fig~\ref{Fig:mrk35sersic} we show three apparently equally good fits: (1),
(2) and (5) from Table~\ref{Tab:fitmrk335}.

\item {\bf Mrk~35} --- Mrk~35 resembles closely Mrk~33 with respect to its LSBC
morphology and the radial extent of its star-forming component ($R \simeq
20$\arcsec). However, at variance to Mrk~33, decreasing the minimum fitting
radius to $R \simeq 12$\arcsec\ has nearly no effect on the S{\'e}rsic exponent
$n$. Fits done in different radial ranges and in the three NIR bands all give
consistent results, with $n \simeq 0.8-0.9$. Thus, the LSBC of Mrk~35 appears
to be well described by a S{\'e}rsic law very close to an exponential,
and whose parameters do not depend significantly on the band or on the
selected radial range. 

\end{itemize}

\subsubsubsection{Sensitivity to sky subtraction errors}

Another important source of uncertainties on the S{\'e}rsic exponent $n$ is
associated with sky-subtraction errors, which can significantly change the SBP
slope in its outermost parts. For an overestimate of the sky-background, the
SBP will curve down, so that the derived $n$ will be lower, and the total
magnitude of the fitted S{\'e}rsic model fainter and for an underestimate of
the sky-background, the profile will  become flatter, so that the fitted $n$
will be higher and the total magnitude brighter.

We checked the importance of this effect on Mrk~33 and Mrk~35. We first
recomputed their $J$ profiles by adding a constant  offset $\pm \Delta {\rm
sky}$ to the input images. The term $\Delta {\rm sky}$ was chosen so as to
change the  intensity of the outermost datapoint of the SBP ($\mu_J=23.68$
\sbb\ at  $R=49.90$\arcsec\ for Mrk 33 and  $\mu_J=23.58$ \sbb\ at
$R=38.64$\arcsec\ for Mrk~35) by 25\% ($\simeq 0.25$ mag). The S{\'e}rsic
parameters fitted to these SBPs are compared with those derived in the
original SBP in Table~\ref{Tab:skymrk335}.

\begin{deluxetable}{lccccccc}
\tabletypesize{\footnotesize}
\tablewidth{0pt}
\tablecaption{S{\'e}rsic parameters for Mrk~33 and Mrk~35 for an 
imperfectly subtracted sky background.\label{Tab:skymrk335}}
\tablehead{\colhead{Profile} & \colhead{Radial range} &
           \colhead{Surface brightness range} & \colhead{$n$} &
           \colhead{rms} & \colhead{$r_e$} & \colhead{$\mu_e$} &
           \colhead{$m_{\rm LSBC}$} \\ 
	   \colhead{}  &      
           \colhead{arcsec}       &
           \colhead{\sbb}                     & \colhead{} &
           \colhead{mag}    &  \colhead{arcsec}  & \colhead{\sbb}  &
           \colhead{mag}
           }
\startdata
\multicolumn{8}{c}{Mrk~33 - $J$ band} \\[2pt]
original              & 20.47--49.90 & 21.23--23.68 & $2.09$ & $0.041$ & 17.76 & 20.99 & 11.68\\
sky under-subtracted  & 20.47--49.90 & 21.20--23.43 & $3.91$ & $0.036$ & 17.23 & 20.87 & 11.32\\
sky over-subtracted   & 20.47--49.90 & 21.26--24.01 & $1.33$ & $0.048$ & 18.15 & 21.08 & 11.95\\[6pt]
\multicolumn{8}{c}{Mrk~35 - $J$ band} \\[2pt]
original              & 15.45--38.64 & 20.32--23.58 & $0.76$ & $0.088$ & 14.61 & 20.23 & 11.84\\
sky under-subtracted  & 15.45--38.64 & 20.31--23.33 & $0.88$ & $0.081$ & 14.51 & 20.19 & 11.74\\
sky over-subtracted   & 15.45--38.64 & 20.33--23.91 & $0.64$ & $0.098$ & 14.79 & 20.28 & 11.94\\
\enddata
\end{deluxetable}

We can see that, when the fit is not well constrained, as is the case for 
Mrk~33, even modest \emph{systematic} uncertainties of $\leq$0.25 mag in the
outer SBP portion can change $n$ by a factor of 2, and model-dependent
estimates of the total magnitude by 0.3 mag. For the case studied here, the 
$r_e$ and $\mu_e$ appear to be more stable. 

Mrk~35, which shows a nearly exponential slope in its LSBC, appears to be more 
robust against sky-subtraction errors with respect to its $n$.

\subsubsection{Can the superposition of two different stellar populations 
mimic a S{\'e}rsic profile?}
\label{sumtwo}

As seen before, good S{\'e}rsic fits can be obtained even when the fitted
radius interval includes a significant portion of the starburst component. So
it is natural wondering whether the sum of two stellar populations differing in
their $M/L$ and radial distribution may reproduce a S{\'e}rsic profile with an
exponent $n$ significantly larger than that of its constituents. If true, this
would imply that a satisfactory S{\'e}rsic fit does not rule out the existence
of two or more distinct major stellar populations, and S{\'e}rsic model
parameters may give a misleading description of the actual photometric
structure of a galaxy. 

This in turn means that one must be careful when using the exponent $n$ 
from a fit to the whole light profile to make a comparative analysis of the
structure of different classes of galaxies or of galaxies at different 
redshifts. 
For instance, assume for sake of discussion $n=3$: this could be just
a normal bulge, or a galaxy actually formed by starburst and a
host component whose cumulative light distribution mimics well a S{\'e}rsic
profile. The presence of two distinct components may go undetected because of
poor resolution, faintness, or just because a very large sample is being
analyzed in some automated way, and no visual inspection of each object is 
done.
A similar effect has been observed by \citet{Balcells03}:
"fits to ground-based profiles reach S{\'e}rsic indices $n > 4$ because the 
light from HST-unresolved central sources, plus in some instances nuclear 
disks or bars, when convolved with typical ground-based seeing, link smoothly
with the extended bulge profile and mimic higher n S{\'e}rsic profiles".

A real-life example is offered by the BCD VII~Zw~403. The ongoing 
star-forming activity in this system is indicated by H$\alpha$ emission, 
traceable in the inner $\ga 0.5$ kpc \citep{Silich02,Papaderos02}. 
Color magnitude diagram {\sl HST} studies \citep{Lynds98,Schulte99} and 
ground-based optical photometry \citep{Papaderos02} show that the young 
blue stellar population is embedded into an old, more extended LSB 
component. 
Deep surface photometry down to $\mu_B \simeq 28$ \sbb\ indicates that the LSB 
emission of VII~Zw~403 can be well approximated by an exponential law 
\citep{Papaderos02}.

Here we have reanalyzed the $B$ band profile derived in this latter study, now
using a S{\'e}rsic model to fit the LSBC of VII~Zw~403. A model with $n=0.97$
yields an excellent fit in the radius range $36\arcsec <R<80\arcsec$; by
subtracting it from the observed profile, we can recover the intensity profile
of the more centrally concentrated starburst population, which in turn is well
fitted by another S{\'e}rsic function with $n=0.86$.  The inner luminosity
component is due to a young stellar population as indicated by its blue
optical colors \citep{Lynds98,Papaderos02}. Hence, the observed SBP of
VII~Zw~403 is well described by the superposition of two nearly exponential
profiles (bottom panel in Fig~\ref{Fig:VIIZw403}), each representing a very
different stellar population with respect to age and $M/L$.

Notwithstanding this fact, if our light profile were truncated at $\mu_B
\la 26.3$ \sbb ($R \simeq50\arcsec$), a single S{\'e}rsic model with $n=3.40$
would successfully fit the observed SBP in the range $5\arcsec <R<50\arcsec$
(upper panel in Fig~\ref{Fig:VIIZw403}).
So, if somebody were dealing with less deep profiles and lacking additional
information on the galaxy, such as color maps or narrow-band images, they
would perhaps conclude that this galaxy has just a single component with
a S{\'e}rsic index $n \sim 3.5$.

This underlines once more the necessity for examining all available
observational evidence (morphology, color profiles, ionized gas distribution, 
etc.) prior to profile fitting and decomposition, as well as the crucial 
importance of deep surface photometry studies for obtaining an unbiased 
description of the underlying host galaxy of BCDs.

\subsection{Deriving the photometric parameters}  
\label{param}

Taking the above remarks into account, we proceeded to fit S{\'e}rsic models
to the SBP of the galaxies in our sample. To minimize the risk that the choice
of the fitted intervals were influenced by a subjective judgment or
expectation on the results and quality of the fit, one of the authors
selected the radial interval upon inspection of the color and \Ha\ maps, and
of the SBP (in order to exclude those outermost datapoints with large error
bars and obviously deviating from the profile's general trend); another author
ran the fitting algorithm within the prescribed radius intervals.

The results are listed in Table~\ref{Tab:Fitresults}; a quick inspection does
confirm that S{\'e}rsic fits to the SBP can in some cases give very different
parameters in the three bands, $n$ varying for a single object by up to 
one dex (Mrk~36). Solutions with S{\'e}rsic exponents $n \gtrsim 15$ are
omitted, because such fits fail to produce meaningful results, a general fact 
already noted by \citet{Graham96}. 

These uncertainties in the derived S{\'e}rsic parameters propagate into large 
uncertainties in the model-dependent total magnitudes ($m_{\rm LSBC}$) 
of the LSBC. As a result, the integral color derived for the underlying host 
galaxy from subtraction of the $m_{\rm LSBC}$ in different bands can in many 
cases be incompatible with the observed ones (those derived directly from
the CP, see next section), and cannot be accounted for by any stellar
population (e.g. Mrk~33 with $(J-H)_{\rm LSB}=-0.23$ together with 
$(J-\ks)_{\rm LSB}$=0.20).

\begin{deluxetable}{llcccccc}
\tabletypesize{\footnotesize}
\tablewidth{0pt}
\tablecaption{S{\'e}rsic photometric parameters for the LSB host galaxy 
\label{Tab:Fitresults}}
\tablehead{\colhead{Galaxy} & \colhead{Band} & \colhead{Radial range} &
           \colhead{Surface brightness range} & \colhead{$n$} & 
	   \colhead{$r_e$} & \colhead{$\mu_e$} & \colhead{$m_{LSBC}$}\\ 
	   \colhead{}  &  \colhead{}  &  \colhead{arcsec}       &
           \colhead{\sbb}   & \colhead{} & \colhead{arcsec}  & 
	   \colhead{\sbb}  &  \colhead{mag} }
\startdata
Mrk~33            & $J$     & 16.89--42.57 & 20.84--23.23 &  2.88 & 15.34    & 20.64 & 11.49 \\
                  & $H$     & 16.26--22.83 & 20.38--21.33 &  0.90 & 13.44    & 20.01 & 11.72 \\
                  & \ks     & 16.09--29.89 & 20.13--21.77 &  1.38 & 15.06    & 20.03 & 11.29 \\
Mrk~35            & $J$     & 21.25--34.53 & 21.12--22.82 &  0.97 & 12.72    & 19.80 & 11.59 \\
                  & $H$     & 21.30--42.88 & 20.49--23.30 &  0.80 & 15.88    & 19.91 & 11.31 \\
                  & \ks     & 21.95--43.70 & 20.19--23.12 &  1.01 & 13.33    & 19.03 & 10.70 \\
Mrk~36            & $J$     & 07.44--11.30 & 21.46--22.77 &  9.46 & \phn0.37 & 13.95 & 12.25 \\
                  & $H$     & 07.55--12.37 & 21.17--22.50 &  0.74 & \phn7.34 & 21.14 & 14.26 \\
                  & \ks     & 07.86--11.34 & 20.98--22.03 &  1.73 & \phn4.67 & 19.78 & 13.46 \\
UM~462            & $J$     & 08.63--21.40 & 20.86--23.12 &  1.20 & \phn8.63 & 20.82 & 13.35 \\
                  & $H$     & 08.04--21.52 & 20.23--22.59 &  1.50 & \phn9.02 & 20.39 & 12.72 \\
                  & \ks     & 08.16--11.73 & 19.98--20.65 &  0.57 & \phn8.90 & 20.12 & 12.94 \\
II~Zw~70          & $J$     & 10.04--25.77 & 21.57--24.76 &  2.30 & \phn5.09 & 19.98 & 13.34 \\
                  & $H$     & 10.17--22.31 & 21.19--23.83 &  2.19 & \phn5.38 & 19.72 & 12.98 \\
                  & \ks     & 09.88--22.52 & 20.88--23.82 &  1.35 & \phn6.57 & 20.07 & 13.14 \\
II~Zw~71          & $J$     & 18.55--27.51 & 21.71--22.94 &  0.70 & 14.77    & 21.27 & 12.89 \\
                  & $H$     & 18.74--27.26 & 21.34--22.71 &  0.34 & 15.36    & 21.02 & 12.89 \\
                  & \ks     & 16.11--27.07 & 20.82--22.36 &  0.46 & 15.41    & 20.79 & 12.52 \\
I~Zw~123          & $J$     & 08.34--11.06 & 21.57--22.46 & $> 15$ & \nodata & \nodata & \nodata  \\
                  & $H$     & 07.68--09.58 & 20.73--21.45 & $> 15$  & \nodata & \nodata & \nodata  \\
                  & \ks     & 06.30--09.17 & 19.95--21.45 &  0.29 & \phn5.11 & 19.67 & 14.01 \\
Mrk~297           & $J$     & 20.42--34.31 & 21.45--24.10 & $> 15$  & \nodata & \nodata & \nodata  \\
                  & $H$     & 20.36--29.29 & 21.29--23.30 &  2.94 &  \phn1.37 & 12.21 & 08.31 \\
                  & \ks     & 18.70--26.84 & 20.41--22.52 & $> 15$ & \nodata & \nodata & \nodata  \\
\enddata
\end{deluxetable}

Due to the failure of the S\'ersic law to produce robust and self-consistent 
models for a large fraction of the available SBPs, it appears worthwhile to 
check whether a fixed $n=1$ (exponential model) allows for more a stable 
approximation to the LSBC.

We start off by repeating the test already done for the S{\'e}rsic law for
Mrk~33, that is, we fit again its $J$ and \ks\ SBP within the same radius
intervals, this time after fixing $n=1$. The results, listed in
Table~\ref{Tab:fitexpmrk33}, show that exponential fits give parameters 
consistent in the different bands, and are not that sensitive to the selected 
radial range (provided it does not include the starburst), nor to 
sky-subtraction errors, as compared with S{\'e}rsic models.

The fitted parameters to the LSBC of all sample galaxies using an exponential 
model are listed in Table~\ref{Tab:Exporesults}. 
At variance to what we found for the S{\'e}rsic models, the integral colors of 
the LSBC, as obtained from exponential fits, are consistent within the errors 
with the colors derived directly from color profiles.  

We note that this improvement in the stability of the fit results is due 
to having set $n$ equal to a constant (thus removing one degree of freedom
in the S{\'e}rsic law), not to some particular property of the exponential 
model. Indeed, fixing any other value within 
reasonable intervals, would give comparable improvements in the fit 
stability.

\begin{deluxetable}{cccccccc}
\tabletypesize{\footnotesize}
\tablewidth{0pt}
\tablecaption{Mrk~33: parameters for the exponential model 
\label{Tab:fitexpmrk33}}
\tablehead{\colhead{Fit ID} & \colhead{Band} & \colhead{Radial range} 
& \colhead{Surface brightness range} & 
\colhead{rms} & \colhead{$\alpha$} & \colhead{$\mu_0$}  & 
\colhead{$m_{\rm LSBC}$}\\[1pt]
 \colhead{}  &      
           \colhead{}      & \colhead{arcsec} &
           \colhead{\sbb}   & \colhead{mag} &
           \colhead{arcsec}    &  \colhead{\sbb}  &
           \colhead{mag}
	   } 
\startdata
(1)  & $J$ &  20.47--49.90 & 21.23--23.68 &  $0.061$  & 13.44 & 19.72 & 12.07 \\
(2)  &     &  10.25--49.90 & 19.71--23.68 &  $0.180$  & 11.16 & 19.12 & 11.87 \\
(3)  &     &  21.29--39.12 & 21.40--22.91 &  $0.040$  & 12.88 & 19.63 & 12.08 \\
(4)  &     &  20.47--39.12 & 21.23--22.91 &  $0.047$  & 12.63 & 19.58 & 12.07 \\
(5)  &     &  10.25--39.12 & 19.71--22.91 &  $0.144$  & 09.98 & 18.89 & 11.89 \\
(6)  & $J$, sky under-subtracted &
              20.47--49.90 & 21.20--23.43 &  $0.069$  & 14.75 & 19.86 & 12.01 \\    
(7)  & $J$, sky over-subtracted  &
              20.47--49.90 & 21.26--24.01 &  $0.053$  & 12.06 & 19.53 & 12.12 \\            
(8)  & \ks &  20.21--34.84 & 20.67--22.28 &  $0.049$  & 10.06 & 18.50 & 11.48 \\
(9)  &     &  10.48--34.84 & 18.98--22.28 &  $0.124$  & 08.65 & 18.02 & 11.32 \\
\enddata
\tablecomments{For the exponential law there is a direct relation between 
the effective parameters $r_e$ and $\mu_e$ and the scale factors 
$\alpha$ and $\mu_0$: $\alpha= 0.596\;R_e$ and $\mu_0=\mu_e-1.822$}
\end{deluxetable}

\begin{deluxetable}{llccccc}
\tabletypesize{\footnotesize}
\tablewidth{0pt}
\tablecaption{Structural parameters of the LSB host galaxy from exponential 
fits. \label{Tab:Exporesults}}
\tablehead{\colhead{Galaxy} & \colhead{Band} & \colhead{Radial range} &
           \colhead{Surface brightness range} & \colhead{$\alpha$} &   \colhead{$\mu_0$} & 
	   \colhead{$m_{\rm LSBC}$}   \\ 
           \colhead{}  &   \colhead{}   & \colhead{arcsec} &
           \colhead{\sbb}   &  \colhead{arcsec}    &  \colhead{\sbb}  &
           \colhead{mag}    }
\startdata
Mrk~33            & $J$     & 16.89--42.57 & 20.84--23.23 &   12.03    & 19.43 & 12.02  \\
                  & $H$     & 16.26--22.83 & 20.38--21.33 &  \phn7.86  & 18.14 & 11.66  \\
                  & \ks     & 16.09--29.89 & 20.13--21.77 &  \phn9.54  & 18.35 & 11.45  \\
Mrk~35            & $J$     & 21.25--34.53 & 21.12--22.82 &  \phn7.45  & 17.93 & 11.56  \\
                  & $H$     & 21.30--42.88 & 20.49--23.30 &  \phn8.34  & 17.71 & 11.10  \\
                  & \ks     & 21.95--43.70 & 20.19--23.12 &  \phn8.05  & 17.26 & 10.72  \\
Mrk~36            & $J$     & 07.44--11.30 & 21.46--22.77 &  \phn3.34  & 19.09 & 14.47  \\
                  & $H$     & 07.55--12.37 & 21.17--22.50 &  \phn4.21  & 19.23 & 14.10  \\
                  & \ks     & 07.86--11.34 & 20.98--22.03 &  \phn3.47  & 18.53 & 13.82  \\
UM~462            & $J$     & 08.63--21.40 & 20.86--23.12 &  \phn5.43  & 19.13 & 13.45  \\
                  & $H$     & 08.04--21.52 & 20.23--22.59 &  \phn5.88  & 18.76 & 12.91  \\
                  & \ks     & 08.16--11.73 & 19.98--20.65 &  \phn5.73  & 18.44 & 12.64  \\
II~Zw~70          & $J$     & 10.04--25.77 & 21.57--24.76 &  \phn5.35  & 19.66 & 14.02  \\
                  & $H$     & 10.17--22.31 & 21.19--23.83 &  \phn5.15  & 19.14 & 13.58  \\
                  & \ks     & 09.88--22.52 & 20.88--23.82 &  \phn4.67  & 18.74 & 13.38  \\
II~Zw~71          & $J$     & 18.55--27.51 & 21.71--22.94 &  \phn7.94  & 19.17 & 12.67  \\
                  & $H$     & 18.74--27.26 & 21.34--22.71 &  \phn6.76  & 18.26 & 12.11  \\
                  & \ks     & 16.11--27.07 & 20.82--22.36 &  \phn8.13  & 18.66 & 12.10  \\
I~Zw~123          & $J$     & 08.34--11.06 & 21.57--22.46 &  \phn3.40  & 18.95 & 14.29  \\
                  & $H$     & 07.68--09.58 & 20.73--21.46 &  \phn2.52  & 17.51 & 13.49  \\
                  & \ks     & 06.30--09.17 & 19.96--21.45 &  \phn2.07  & 16.58 & 12.99  \\
Mrk~297           & $J$     & 20.42--34.31 & 21.45--24.10 &  \phn5.87  & 17.90 & 12.05  \\
                  & $H$     & 20.36--29.29 & 21.29--23.30 &  \phn4.86  & 16.80 & 11.36  \\
                  & \ks     & 18.70--26.84 & 20.41--22.52 &  \phn4.20  & 15.67 & 10.54  \\
\enddata
\end{deluxetable}

\subsection{Colors of the underlying stellar host galaxies and age estimation} 
\label{colorhost}

Upon inspection of the CPs in Figure~\ref{Fig:SBP1} we see 
that the galaxies present constant colors in their outer regions, in 
agreement with the results found in the optical (Papers~I and II). The good
quality of our data allowed us to derive reliable colors for the LSBC of
the BCDs. They have been computed by averaging the color profiles over a radial
range which excludes the starburst-dominated regions (i.e., for 
$R\geq R_{\rm transition}$) and those outermost datapoints having large 
error bars. The resulting values, corrected from Galactic extinction as
described in Section~\ref{isomag}, are displayed in Table~\ref{Tab:LsbPhot}.

To obtain an estimate of the ages of LSBCs, we compared their colors  with the
predictions of the GALEV spectrophotometric evolutionary synthesis models
\citep[][ Anders, Bicker \& Fritze-v.Alvensleben, priv. comm.]{Schulz02}. 
An initial mass function with a Salpeter slope and lower and upper cutoff 
masses of 0.08 and 100 M$_{\odot}$ was chosen.

Figure~\ref{Fig:modelos} displays the (\vj)--(\jh) diagram for the LSBC of
the sample galaxies, together with the tracks of the models for an
instantaneous burst (IB) with metallicities $Z=0.02\,Z_\odot$ and 
$Z=0.4\,Z_\odot$, as well as for a continuous star-formation (CSF) rate 
with the same metallicities. 
Most of the galaxies seem to present colors compatible with the models 
predictions. In the IB approximation, the ages of the LSBC are estimated to be 
$\ga 3$ Gyr, whereas, as expected, ages are larger in the CSF approximation.

Two of the objects, Mrk~33 and Mrk~297, which happen to be the most luminous 
galaxies in the sample, lie far off the model tracks. 
A detailed study of the evolutionary status and stellar composition of these 
BCDs is outside the scope of the present paper. In fact a proper 
characterization of their stellar population, done by combining multiband
images and spectra, will be the subject of the next paper in this series.

\begin{deluxetable}{lccccccc}
\tabletypesize{\footnotesize}
\tablewidth{0pt}
\tablecaption{Colors of the LSB host galaxy  \label{Tab:LsbPhot}}
\tablehead{\colhead{Galaxy} &
              \colhead{\vj} & \colhead{\vh} & \colhead{\vk} &
	   \colhead{\jh} & \colhead{\hk} & \colhead{\jk} 
           }
\startdata
Mrk~33       & $2.01\pm0.04$ & $2.25\pm0.09$ & $2.52\pm0.05$ & $0.34\pm0.07$ & $0.27\pm0.08 $& $0.51\pm0.05$ \\
Mrk~35       & $1.55\pm0.06$ & $2.17\pm0.05$ & $2.44\pm0.05$ & $0.57\pm0.08$ & $0.30\pm0.05$ & $0.87\pm0.06$  \\
Mrk~36       & $1.35\pm0.12$ & $1.90\pm0.11$ & $2.00\pm0.07$ & $0.45\pm0.12$ & $0.19\pm0.10$ & $0.64\pm0.08$ \\
UM~462       & $1.41\pm0.09$ & $1.97\pm0.19$ & \nodata       & $0.57\pm0.13$ & $0.19\pm0.13$ & $0.77\pm0.03$  \\
II~Zw~70     & $1.49\pm0.07$ & $1.90\pm0.08$ & $2.02\pm0.13$ & $0.40\pm0.04$ & \nodata       & $0.55\pm0.08$  \\
II~Zw~71     & $1.57\pm0.09$ & $1.91\pm0.09$ & $2.16\pm0.08$ & $0.38\pm0.03$ & $0.18\pm0.02$ & $0.57\pm0.04$ \\
I~Zw~123     & $1.43\pm0.03$ & \nodata       & \nodata       & $0.46\pm0.12$ & \nodata       & $0.78\pm0.09$  \\
Mrk~297      & $1.57\pm0.19$ & $1.81\pm0.10$ & $2.03\pm0.08$ & $0.18\pm0.08$ &$0.31\pm0.05$  & $0.51\pm0.06$  \\
\enddata
\tablecomments{Columns 2-7: mean optical-NIR and NIR colors, corrected
for Galactic extinction, of the LSBC of the sample galaxies, as derived 
from their color profiles.}
\end{deluxetable}

\section{Comments on the Individual Galaxies}
\label{objects}

{\it Mrk~33}. --- Mrk~33 is a luminous BCD with a regular appearance,
belonging to the {\it Type~I\/} (nuclear starburst) BCD class introduced in
Paper~II; the starburst activity is confined to the inner regions of a smooth,
red stellar LSBC. It is a well studied galaxy, with a metallicity of about
$1/3\;Z_\odot$ \citep{davidge89,Legrand97}. \citet{Kunth85} reported the
presence of a large number of Wolf-Rayet (WR) stars in its center, while
\citet{Lequeux95} and \citet{Legrand97} reported evidence for a strong 
outflow of gas, ejected from the starburst region at velocities close to 200 km
sec$^{-1}$. To study its stellar population, \citet{fanelli88} analyzed the UV
spectrum and concluded that the present burst of SF has been preceded by at
least two other episodes, the most recent of which ended not more than 20 Myr
ago. \citet{kruger95} obtained also a good fit to the optical and NIR continuum
by assuming that a 5 Myr long starburst took place in a galaxy which had
been forming stars continuously, but at a lower rate, during the last 15 Gyr.
\citet{Mashesse99} estimated an age of 4.5 Myr for the present starburst. 

The morphologies in the NIR and in the optical basically coincide, though in
the optical the inner isophotes, clearly affected by the presence of dust, have
a much more irregular appearance than in the NIR. Also, in the NIR we can
better distinguish two peaks in the central starburst, aligned in the 
southeast-northwest direction; their positions coincide with the knots visible
in the optical color maps (see Paper~II).

The \vk\ color morphology of Mrk~33 (Figure~\ref{Fig:VKmkn33colormaps}) is
markedly different from that seen on optical color maps (see Paper~II).  An
inhomogeneous extinction pattern in the inner portions of the galaxy, already
noticed in the optical (\citealt{Mollen92}; Paper~II), is confirmed by our
color maps: a conspicuous red patch is seen at the south-east of the galaxy
center. Interestingly, the bluest regions do not coincide with the brightest
(central) knot, as it was found to be the case in optical color maps, but
appear displaced to the south.

We have performed aperture photometry centered on the red peak, within a
circular aperture of radius 2\arcsec, and determined its colors: $\vj=2.18$,
$\vh = 2.74$, $\vk = 3.00$, $\jh = 0.56$, $\hk = 0.26$. Such optical-NIR
colors, which are even redder than those in the LSB component (see
Table~\ref{Tab:LsbPhot}), can naturally be explained by dust extinction. This
is also supported by recent interferometric $^{12}$CO studies of Mrk~33
revealing a morphologically and kinematically complex molecular gas
distribution in its inner part, also pointing to a high, non-uniform absorption
\citep{Fritz00}.

{\it Mrk~35}. ---  This is another luminous, high-metallicity BCD, with an
oxygen abundance $Z_\odot/3$ \citep{Steel96}, which belongs to the  {\it
Type~III\/} (chain starburst) morphological class introduced in Paper~II.  The
star-forming activity is distributed along the northeast-southwest direction,
in a bar-like structure. The brighter star-forming regions are placed in the
central part of the galaxy, and form a "heart-shaped" structure; a tail departs
from it to the south-west, connecting with two moderately bright knots 
$\sim$20\arcsec\ further away.  We have labeled the knots resolved in the
broad-band frames following \citet{MazBor93} (see Figure~\ref{Fig:SBP3}).
\citet{Steel96} reported the presence of Wolf-Rayet stars in the knot {\sf a},
and inferred for it an age younger than 5 Myr. 

The morphology in the optical and NIR broad-band frames is similar. Optical
frames present more distorted inner isophotes, which point to the presence of
dust. Interestingly, the central knot, {\sf b}, coincides with the intensity
peak in the NIR, whereas knot {\sf a} is identifiable with the optical maximum.
The intensity ratio between knots {\sf a} and {\sf b} decreases when moving
from the $U$ to the \ks\ band. A similar behavior has been reported so far only
in the BCD Mrk~178 by \cite{Noeske03}.

Figure~\ref{Fig:VKmkn35colormaps} displays the \vk\ color map of Mrk~35, which
reveals a very intriguing morphology: a dust lane is crossing the center of the
galaxy in the north-south direction, bending eastwards to the south. The {\sf
a} knot has blue colors, while the inner knots {\sf b} and {\sf c} are very
red.

{\it Mrk~36}. --- It is a typical compact, low-luminosity and low-metallicity
BCD, classified as {\it Type~I\/} (nuclear starburst).  Star-forming activity
takes place in the inner regions of a roughly elliptical LSBC with redder
colors. It have been shown (Paper~II), from high-resolution ground-based
optical imagery, that star formation occurs in at least five compact 
(diameter$\la 200$ pc) Super-Star Cluster (SSC) candidates spread  over the
inner region of the galaxy. This finding is confirmed by our infrared data. The
ages of these star-forming knots, derived from optical colors, range between
2.5 and 5 Myr \citep{Mashesse99,Cairos00}.

{\it UM~462}. --- This is another low-luminosity, low-metallicity system, with
$12+\log(O/H)=7.95\pm0.01$ \citep{Izotov99}. It falls into the {\it Type~II} 
(extended starburst) morphological class. It exhibits an intriguing morphology
in its star-forming region, which is resolved into six knots. These features
were not detected in our previous optical data (Paper~I) due to poor seeing
conditions. 

\cite{Vanzi02} have recently published NIR spectra and images of this galaxy.
In Figure~\ref{Fig:SBP4} we labeled the star-forming knots following
the notation used by the above authors, who identified them with SSCs. 
Photometry within apertures of
2\arcsec\ in radius is listed in Table~\ref{Tab:knotsum462}. These values have
been corrected only for Galactic extinction; no corrections have been made
for the ionized gas contribution nor for the contribution of the underlying
LSBC. Our results are in fair agreement with the values published by
\cite{Vanzi02}.

\begin{deluxetable}{lcc}
\tabletypesize{\footnotesize}
\tablewidth{0pt}
\tablecaption{Colors of star-forming knots in
UM~462\label{Tab:knotsum462}}
\tablehead{\colhead{Knot} & \colhead{\jh} & \colhead{\hk}\\[1pt]
 \colhead{} & \colhead{(mag)} & \colhead{(mag)}}
\startdata
1 &   0.34    & 0.43  \\
2 &   0.44    & 0.29  \\
3 &   0.39    & 0.35  \\
4 &   0.43    & 0.29  \\
5 &   0.48    & 0.25  \\
6 &   0.43    & 0.29  \\
\enddata
\end{deluxetable}

{\it II~Zw~70}. --- This is a {\it Type~I\/} (nuclear starburst) galaxy, which
forms together with II~Zw~71 a pair of interacting galaxies 
\citep{Vorontsov77}. Evidence for an ongoing interaction between both systems
is presented by interferometric \ion{H}{1} studies \citep{Balkowski78,Cox01}. A
conspicuous central star-forming region is placed in the center of a markedly
elongated LSBC, with a nearly constant ellipticity of $\approx$ 0.7.  An
extended, extraordinarily faint ($\mu_{B} \ga 27$ \sbb) stellar plume emerging
from the star-forming region and protruding $\sim$1\farcm 7 to the direction of
II~Zw~71 has been recently discovered on deep optical images
\citep{Papaderos01,Cairos03}. It apparently coincides with the massive
\ion{H}{1} stream connecting II~Zw~70 with II~Zw~71; its relatively blue 
optical colors are consistent with the hypothesis that it has recently  formed
within the gaseous bridge between the BCDs. The \vk\ color map of II~Zw~70 is
shown in Figure~\ref{Fig:VKiizw70colormaps}.

{\it II~Zw~71}. ---This galaxy belongs to the {\it Type III\/} (chain
starburst) morphological class. 
The interpretation of II~Zw~71 as a (possible) polar ring galaxy 
\citep{Whitmore90} has recently received further observational support through 
the analysis by \citet{Cox01}. 
Its LSBC presents red colors and elliptical isophotes, however the chain of 
star-forming regions, aligned perpendicular to the major axis of the LSBC,
results in an overall distorted morphology. The \bk\ color map of this system 
is shown in Figure~\ref{Fig:VKiizw71colormaps}.

{\it I~Zw~123}. --- It is a small and compact {\it Type~I\/} class BCD, with 
a metallicity of $Z_{\odot}/7$ \citep{Izotov97}.  It presents a regular
morphology, with its star formation occurring atop the central part of a
roughly circular LSB envelope, whose ellipticity is about 0.1. This central
star-forming region is resolved  into two smaller knots. Knot {\sf a} is slightly
displaced from the center of the outer isophotes and shows a slight north-west
elongation; {\sf a} seems to be connected by a "bridge-structure" with the
southern knot {\sf b}. Feather-like extensions can be seen to the northeast and
southwest; they show very red colors, and are not detected in \Ha\ frames
\citep{Noeske01}, so they are most likely background objects. 

{\it Mrk~297}. --- This object ({\it Type II}: extended starburst), originally
discussed in the \citet{Thuan81} list and included in several subsequent
studies of BCDs, strictly speaking should not be regarded as a dwarf galaxy
($M_{B}=-20.54$, Paper~II). \citet{Alloin79} classified it
as a merging system. Indeed, the galaxy is very distorted and shows in its 
LSBC filamentary north-south extensions, suggestive of an ongoing merger
event.

Mrk 297 presents a very intriguing morphology. Its intensity distribution
peaks in the center of the galaxy. Smaller knots form a chain
aligned in the east-west direction. The \Ha\ maps (see Paper~II) show that 
the SF is taking place in numerous regions spread over nearly the entire 
galaxy.
Ground-based \vk\ (see Figure~\ref{Fig:VKmkn297colormaps}) and HST \vr\ color
maps \citep{Papaderos98b} clearly indicate a large-scale, inhomogeneous
absorption pattern in the galaxy .

{\it Mrk~407}. --- It is a very compact and faint {\it Type~I\/} BCD. It
presents an overall regular morphology, though its outer isophotes are slightly
perturbed and elongated along the northwest--southeast direction. The quality
of the NIR images does not permit to tell whether or not an older stellar LSBC 
is underlying its star-forming regions.

\section{Discussion}
\label{discuss}

From the results of the fits listed in Table~\ref{Tab:Fitresults}, it is
evident that the S{\'e}rsic law, while allowing for a more precise description
of the intensity distribution of many LSBCs, still has serious drawbacks, such 
as the sensitivity of the model parameters to the fitted radial range.
Furthermore, in several cases the derived S{\'e}rsic indices for the same
galaxy may be quite different in various passbands. Given that color gradients 
within the LSBC of evolved BCDS are small, we would expect to obtain similar 
S{\'e}rsic solutions regardless of the passband. 

Only in two of the nine sample BCDs, namely Mrk~35 and II~Zw~70, the fit is 
reasonably good and stable. For UM~462 the fits in $J$ and $H$ are in 
agreement; the discrepant parameters obtained for \ks\ are probably due to the 
poorer quality of the data.  

What does all this mean? Should we principally discard the S{\'e}rsic law as a
physically meaningful representation of the SBPs of the host galaxy of BCDs,
and hence look for alternative fitting formulas? Or, more simply, are
observational uncertainties/limitations of the data responsible for the failure
of the S{\'e}rsic law to provide an effective description of the photometric
structure of the LSBC? At this stage an answer to the above questions is
premature. A more comprehensive analysis must be carried out on accurate SBPs
derived from deep multiwavelength data; this will be done in a forthcoming
paper.

The first and immediate conclusion we can draw from this analysis is that one
should exercise extreme caution when fitting the S{\'e}rsic law in BCDs and,
even more importantly, when fitted parameters are discussed in the context 
of the physical state of the stellar LSBC component or the dark matter 
distribution of a BCD. 
We suspect that the problems so far discussed are the reason for the 
surprisingly high S\'ersic exponents derived by \citet{Bergvall02} for the 
LSBC of BCDs: in four of the six galaxies they studied they infer a $n > 17$ 
(see their Table 5). 
It is hardly imaginable that such values, exceeding even those found 
among the brightest elliptical galaxies \citep{Graham96}, are witnessing 
a particular type of dynamical equilibrium of the visible and dark matter in 
BCDs (see discussion in \citealt{Bergvall02}).

We would like to highlight some results that have emerged from the present 
study. First, in some galaxies the profile of the LSBC presents a clear 
curvature in the $R-\mu$ diagram (Mrk~33 and II~Zw~70), and the exponential law
gives a poor fit to their SBP; in other BCDs (Mrk~35 and UM~462), however, the
LSBC seems to be well fit by the exponential law.  This variation in the
profile shape might indicate that the structure of the  LSBC is not the same in
all BCDs, a result already suggested by the findings of previous work
\citep{Loose86,Kunth88,Kunth00,Noeske00,Cairos00,Cairos02,Bergvall02}.

Therefore, the "so called" BCD are objects with an active starburst
which dominates their optical properties, but when the characteristic of LSBC
are taken into account, they seems to constitute an heterogeneous class of
galaxies. 

Second, it is clear that an incorrect modeling of the underlying stellar
component would yield a misleading physical description not only for the  LSB
component itself, but also for the superposed starburst component. We see 
comparing the results from Tables~\ref{Tab:Fitresults}
and~\ref{Tab:Exporesults} that fitting different models to the LSBC one obtains
total LSBC luminosities differing by several tenths of a magnitude (see for
instance Mrk~33 in $J$). Therefore, as pointed out in \citet{Cairos02} and
\citet{Papaderos02}, a proper modeling of the radial intensity distribution of
the LSBC is essential  for an accurate derivation of the star formation history
within the starburst component.

In the 8 galaxies in which a LSBC has been identified, the colors of the latter
suggest old ages. However, two of the galaxies, Mrk~33 and Mrk~297, show
colors which place them in a different part of the two-color diagram (Fig.
4), far from the region populated by other BCDs and described by simple star
formation histories. These two galaxies are the two most luminous in the
sample; according to a strict classification criterion, they should not be
considered dwarf, but belong to the  class of Luminous Blue Compact Galaxies
(LBCG) \citep[][ Paper~II]{Oestlin98}. 

Finally, the fact that we found inhomogeneous, large-scale dust absorption in
three of the nine galaxies challenges the belief that dust is generally of no 
concern in BCD studies. 
This should be taken into account, especially when interpreting ages of BCDs 
using color-magnitude diagrams.

\section{Summary and Conclusions}

This is the third of a series of papers devoted to a multiwavelength study of a
large sample of BCD galaxies. In the first two papers we presented optical
broad- and narrow-band observations of the galaxy sample. Here we have extended
our dataset with broad-band $J$, $H$ and $K_s$ NIR images for 9 BCDs. We have
derived contour maps in the $J$ band, as well as surface brightness and color
profiles. We have also shown optical-NIR color maps for several objects.

The main results of our work can be summarized as follows:
\begin{itemize} 

\item The galaxies present very similar morphologies in the NIR and in the
optical. 
The presence of two distinct components, already seen in previous optical 
studies, is also discernible in the NIR surface brightness 
profiles in all but one case (Mrk~407):
a young starburst population, dominating the inner regions, is superposed
on the host galaxy, detectable at faint surface brightness levels (typically 
$\mu \ga 21$ \sbb\ in $J$). The host galaxy exhibits in its outer regions quite 
regular elliptical isophotes and nearly constant red colors, indicative of an 
evolved stellar population.  
In Mrk~407, the present data do not allow to verify the presence of an old 
underlying low-surface brightness component (LSBC).

\item We have checked whether a S\'ersic fit to the outer part of SBPs can be 
used to parametrize the structural properties of the underlying LSBC. At least 
two of the sample BCDs (namely, Mrk~33 and II~Zw~70), show a clear curvature in
their profiles in the  $R-\mu$ diagram, suggesting a best-fitting S\'ersic
exponent $n>1$, whereas in two other galaxies (Mrk~35 and UM~462) the LSBC
profiles are consistent with an exponential law ($n=1$). We have discussed at
length the problems emerging from profile fitting using  the S{\'e}rsic law,
and shown that the usable surface brightness interval of  the available SBPs
does generally not allow to tightly constrain the  S{\'e}rsic parameters. The
latter were found to be very sensitive to the  fitted radial range and to
uncertainties in the sky subtraction. 

In addition we found that a good S{\'e}rsic profile is not an unambiguous
proof of a monocomponent stellar system; we demonstrate that such a profile
can be produced by the superposition of a young stellar component on an old,
more extended LSBC with an exponential profile.  We emphasize that one must
not underestimate the various sources of  uncertainties affecting S{\'e}rsic
fits to composite and in many respects  very complex extragalactic systems,
such as BCDs. We therefore recommend against just using the apparently
excellent quality of S{\'e}rsic fits to a wide range of observed surface
brightness distributions to derive astrophysical results or inferences.

We have argued that an exponential model, while it may not be a universal 
descriptor of the intensity distribution of the LSBC, it nonetheless appears 
to be considerably more stable against the above mentioned problems. 
Therefore it can provide, at this stage, a more reliable tool to quantify the 
structural parameters of the LSBC and compare them with those of other nearby 
dwarf classes, as well as with star-forming galaxies at a larger redshift.

\item Optical-NIR color maps indicate a large-scale inhomogeneous absorption
in three of the sample galaxies. 

\end{itemize}

\acknowledgments

Based on observations with the WHT operated on the island of La Palma by the 
Royal Greenwich Observatory in the Spanish Observatorio del Roque de los 
Muchachos of the Instituto de Astrof\'\i sica de Canarias and on observations 
taken at the German-Spanish Astronomical Center, Calar Alto, Spain, operated
by the Max-Planck-Institut f{\" u}r Astronomie (MPIA), Heidelberg, jointly
with the spanish "Comision Nacional de Astronomia".  We thank the staff of both
observatories. We thank Dr. A.W. Graham for his valuable comments and  
discussions.
We thank J.~N. Gonz\'alez-P\'erez for his help in the initial stages of this 
project. We thank Dr. Uta Fritze-v.Alvensleben, P. Anders, J. Bicker and
J.Schulz for kindly providing the GALEV models. This work has been partially
funded by the spanish ``Ministerio de Ciencia y Tecnologia'' (grants
AYA2001-3939  and PB97-0158). L.~M. Cair\'os acknowledges support by the EC
grant HPMF-CT-2000-00774. Research by K.G.N. has been supported by the Deutsche
Forschungsgemeinschaft (DFG) grant FR325/50-1. N. Caon thanks the
Universit{\"a}ts-Sternwarte of G{\"o}ttingen for its kind  hospitality. This
research has made use of the NASA/IPAC Extragalactic Database (NED),  which is
operated by the Jet Propulsion Laboratory, Caltech, under contract  with the
National Aeronautics and Space Administration.

\clearpage

\begin{figure*}   
\includegraphics[angle=270,width=\textwidth]{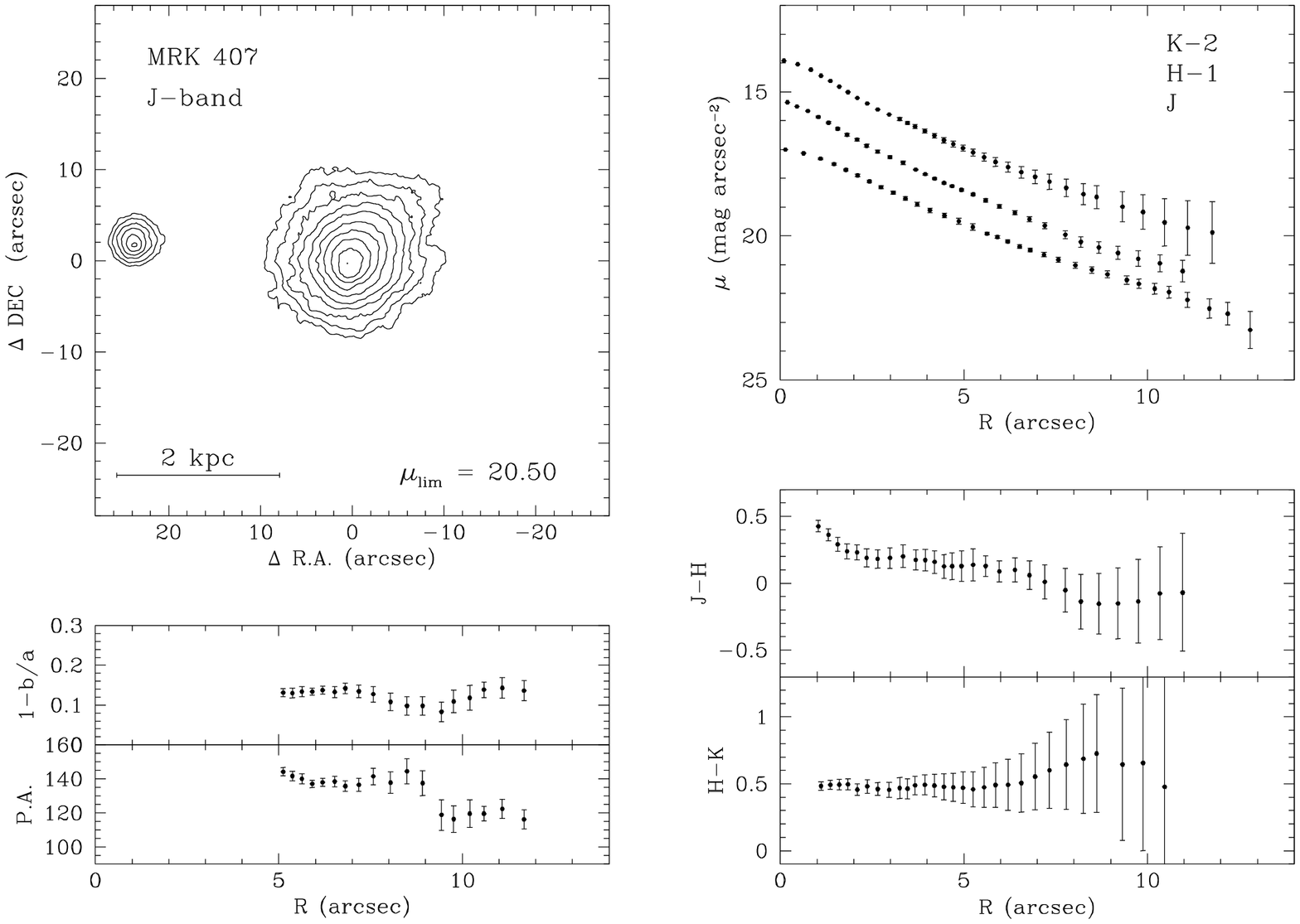}
\caption{Top-left panel: Contour map of the galaxy. The contours
are spaced by 0.5 mag.
The surface brightness level corresponding to the outermost contour and the 
spatial scale are shown in each panel. North is up and East to the left. 
Top-right panel: Surface brightness profiles; $R$ is the equivalent
radius.
Bottom-left panel: Position angle and ellipticity profiles (empty circles 
mean that the ellipse fitting algorithm fails to compute the isophote's P.A. 
and ellipticity, and adopts instead the last measured value). 
Bottom-right panel: Color profiles.}
\label{Fig:SBP1}
\end{figure*}

\addtocounter{figure}{-1}
\begin{figure*}
\includegraphics[angle=270,width=\textwidth]{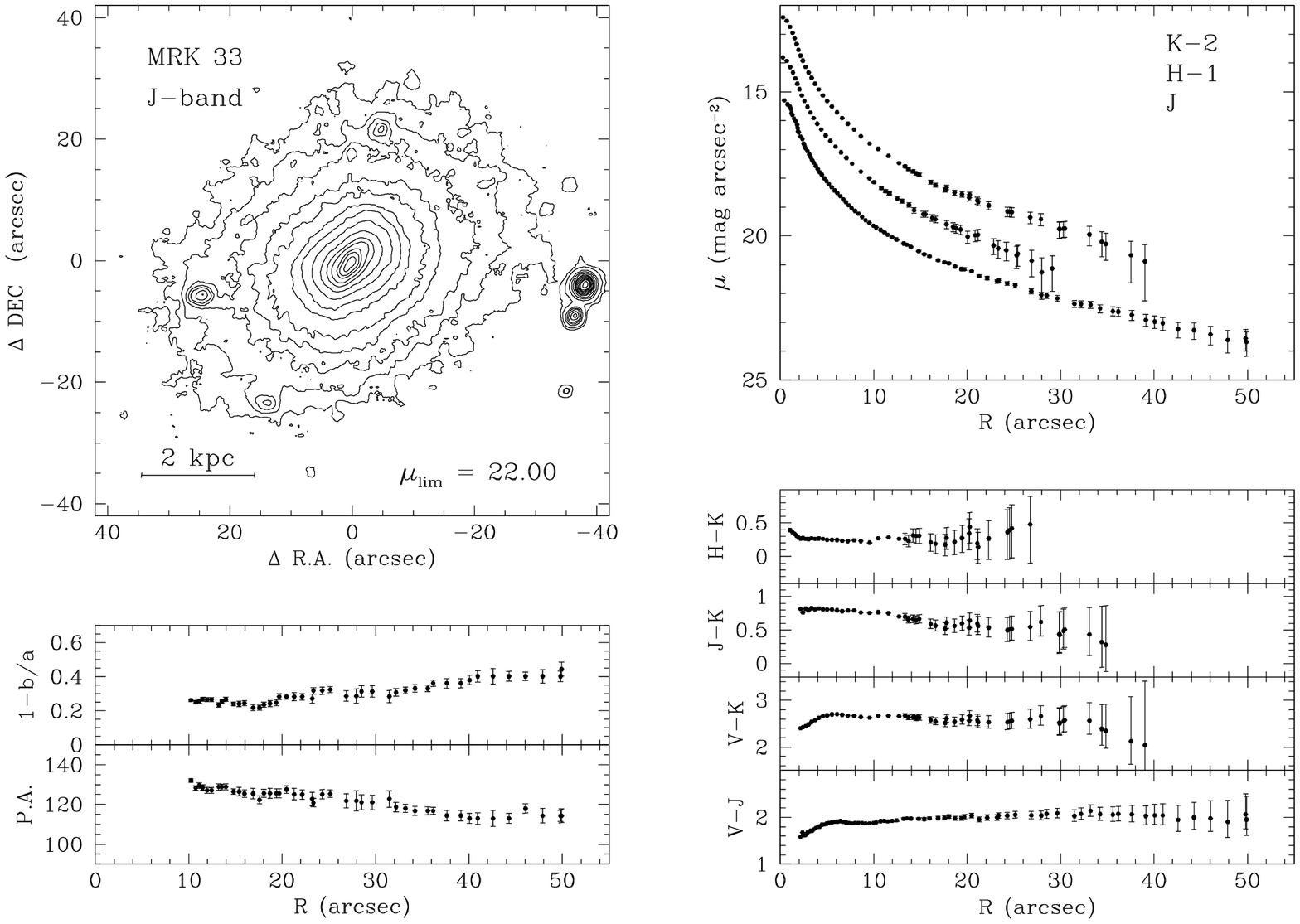}
\caption{\mbox{Continued}}
\label{Fig:SBP2}
\end{figure*}

\addtocounter{figure}{-1}
\begin{figure*}
\includegraphics[angle=270,width=\textwidth]{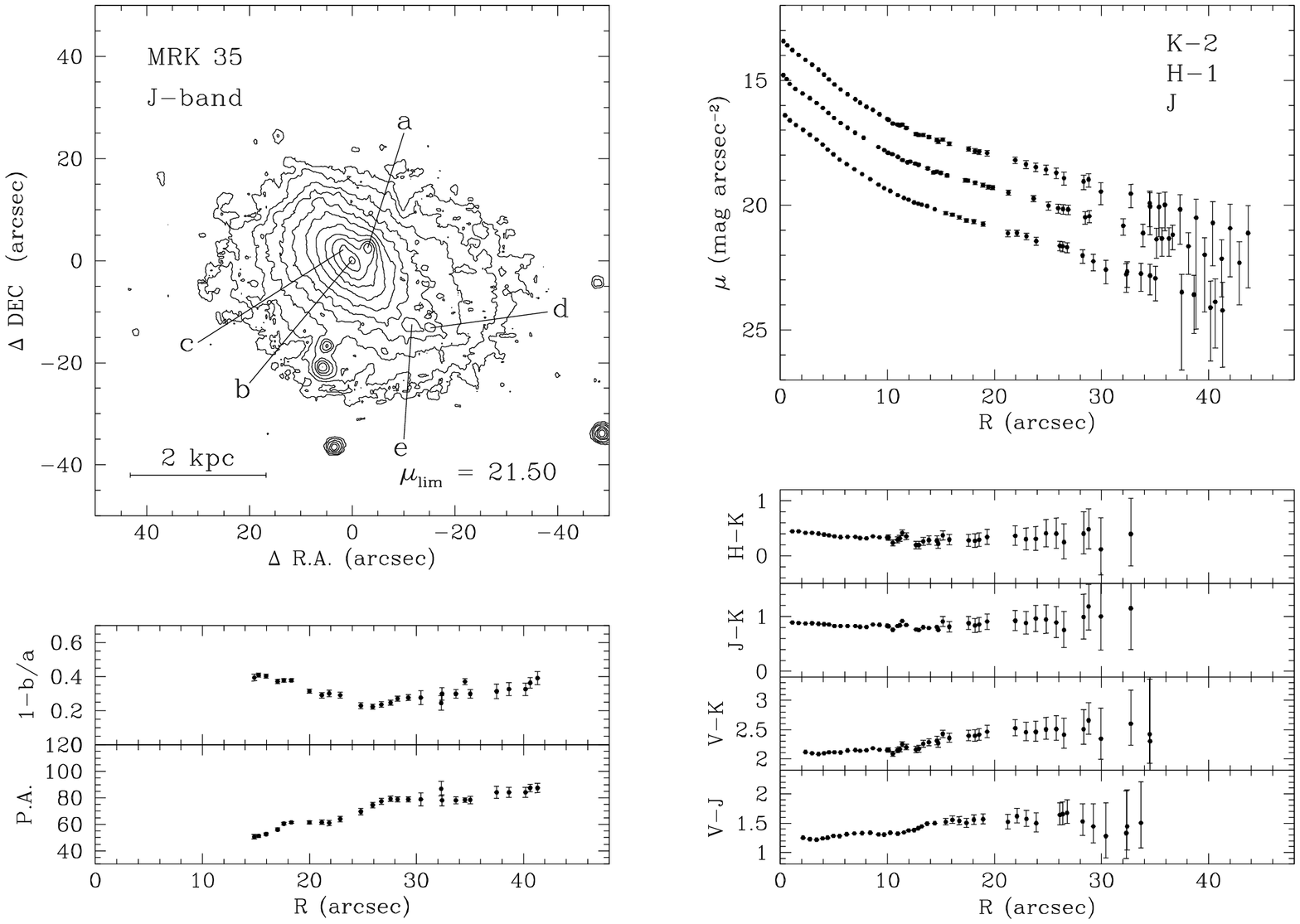}
\caption{\mbox{Continued}}
\label{Fig:SBP3}
\end{figure*}

\addtocounter{figure}{-1}
\begin{figure*}
\includegraphics[angle=270,width=\textwidth]{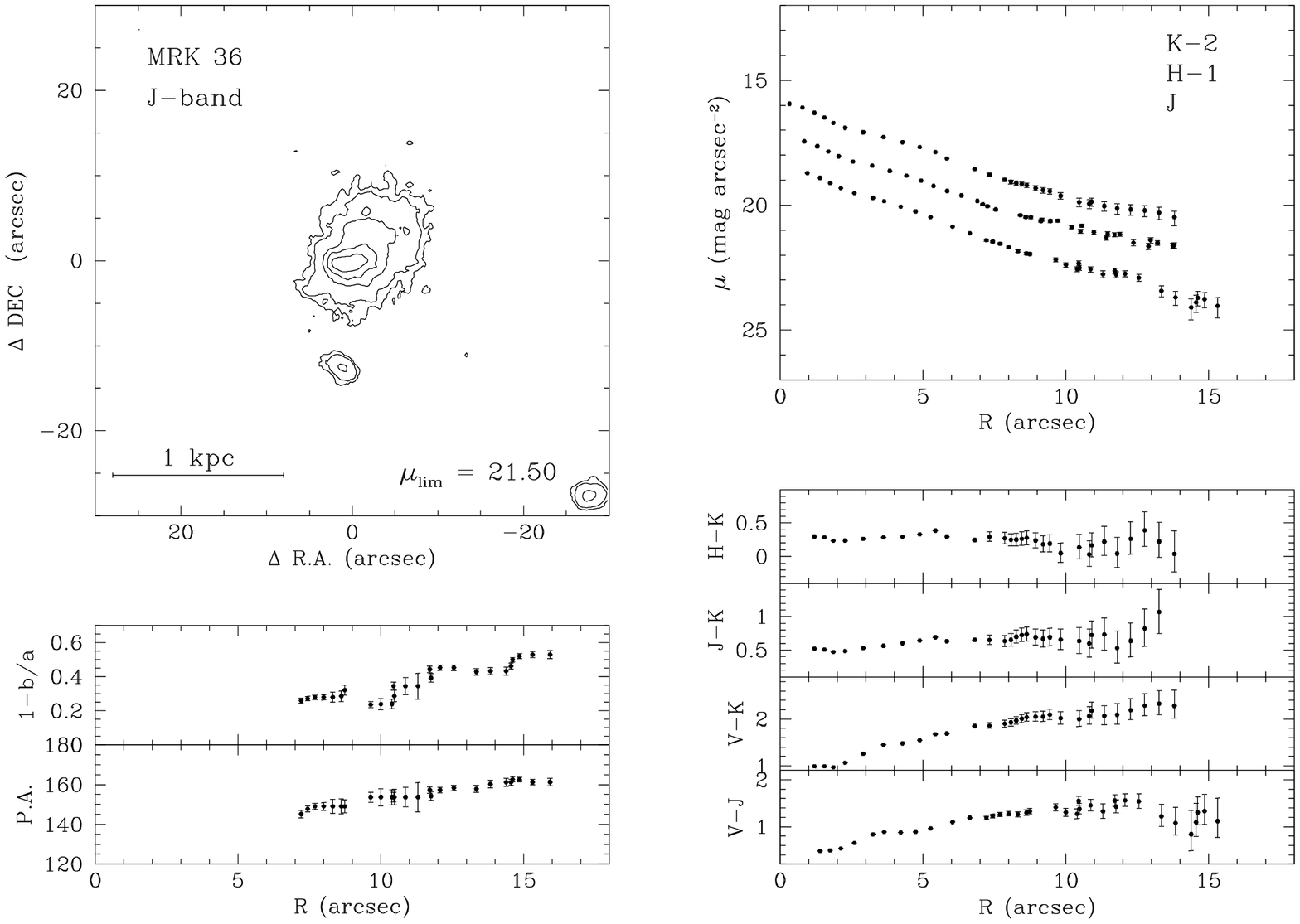}
\caption{\mbox{Continued}}
\label{Fig:SBP4}
\end{figure*}

\addtocounter{figure}{-1}
\begin{figure*}
\includegraphics[angle=270,width=\textwidth]{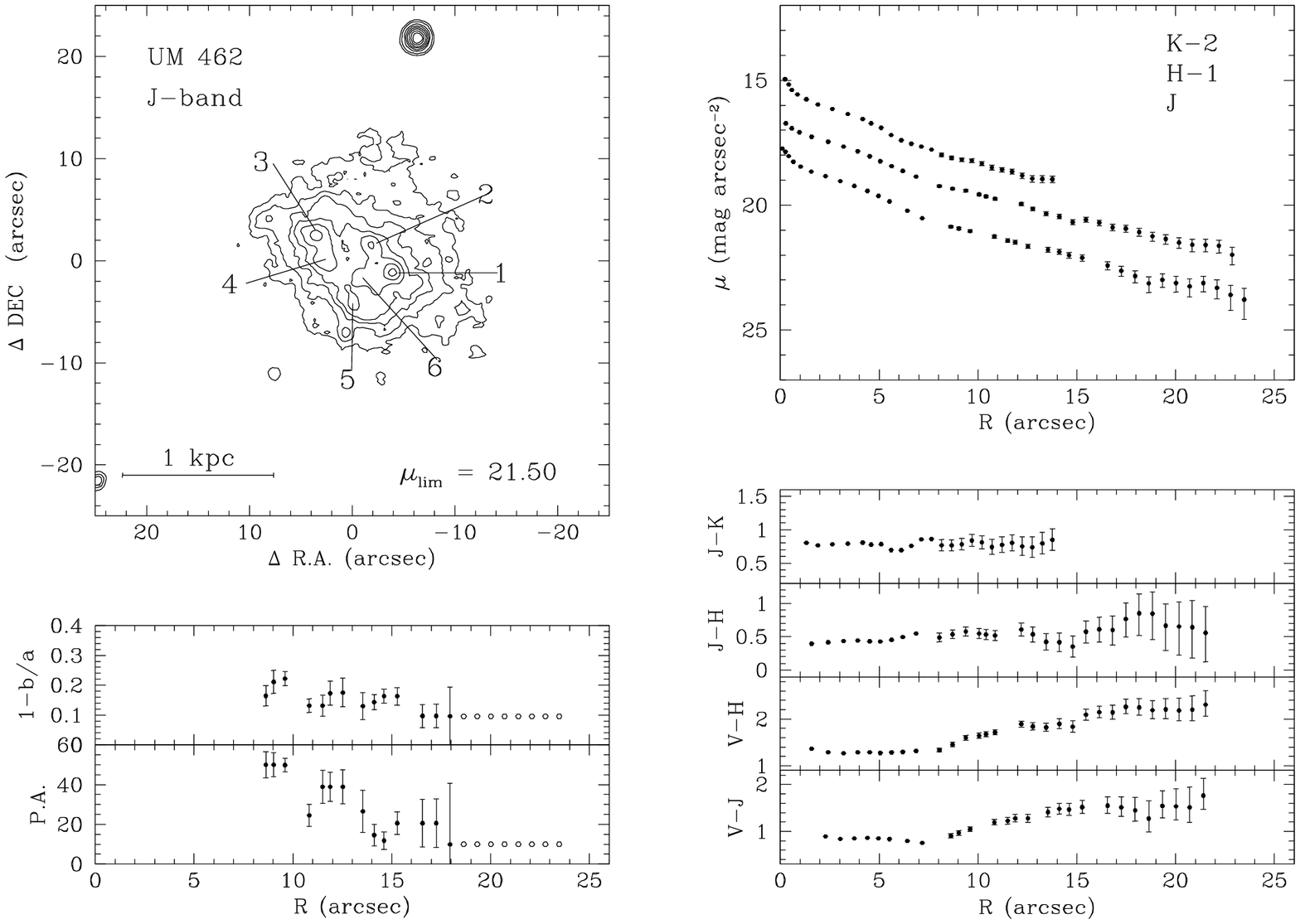}
\caption{\mbox{Continued}}
\label{Fig:SBP5}
\end{figure*}

\addtocounter{figure}{-1}
\begin{figure*}
\includegraphics[angle=270,width=\textwidth]{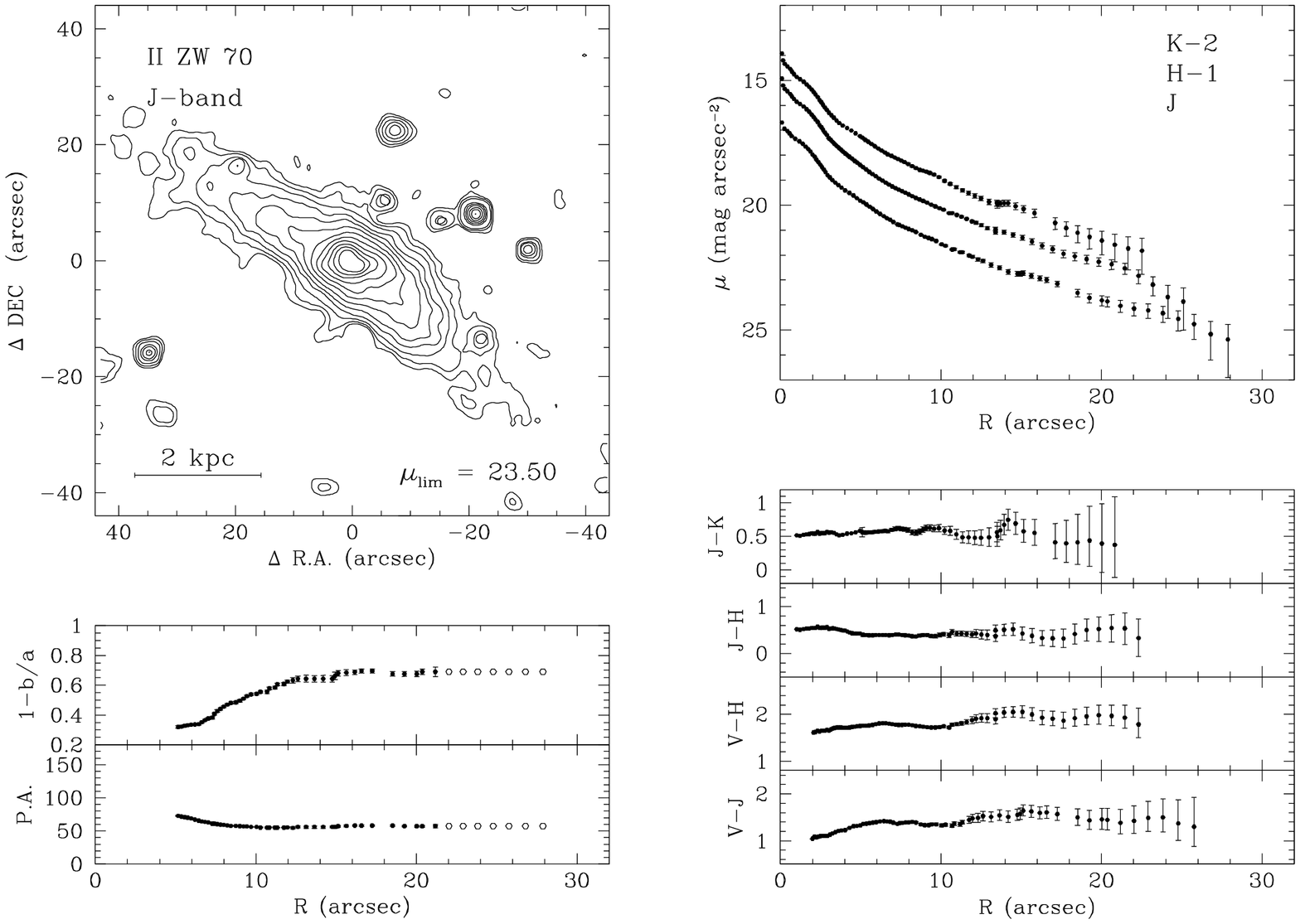}
\caption{\mbox{Continued}}
\label{Fig:SBP6}
\end{figure*}

\addtocounter{figure}{-1}
\begin{figure*}
\includegraphics[angle=270,width=\textwidth]{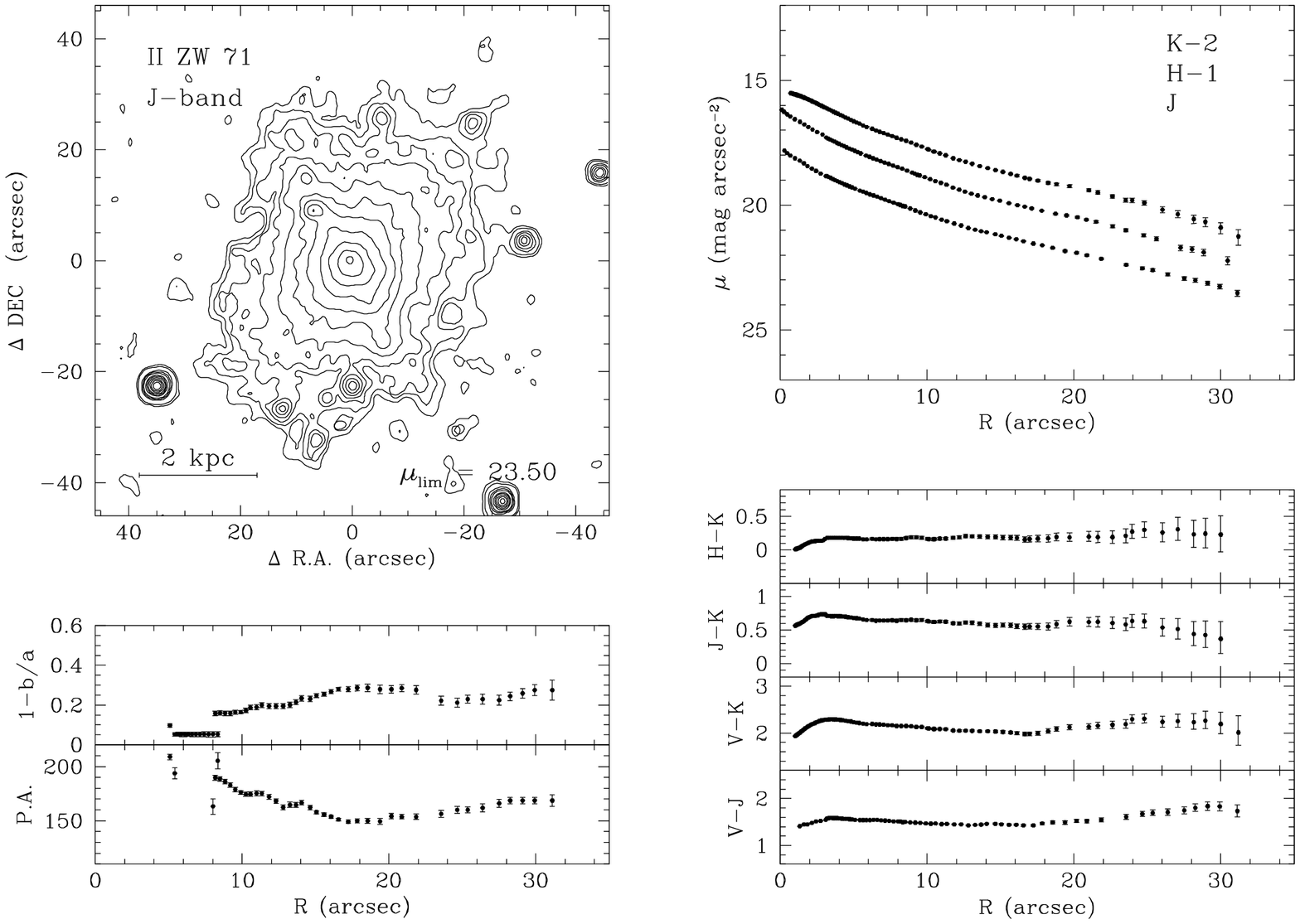}
\caption{\mbox{Continued}}
\label{Fig:SBP7}
\end{figure*}

\addtocounter{figure}{-1}
\begin{figure*}
\includegraphics[angle=270,width=\textwidth]{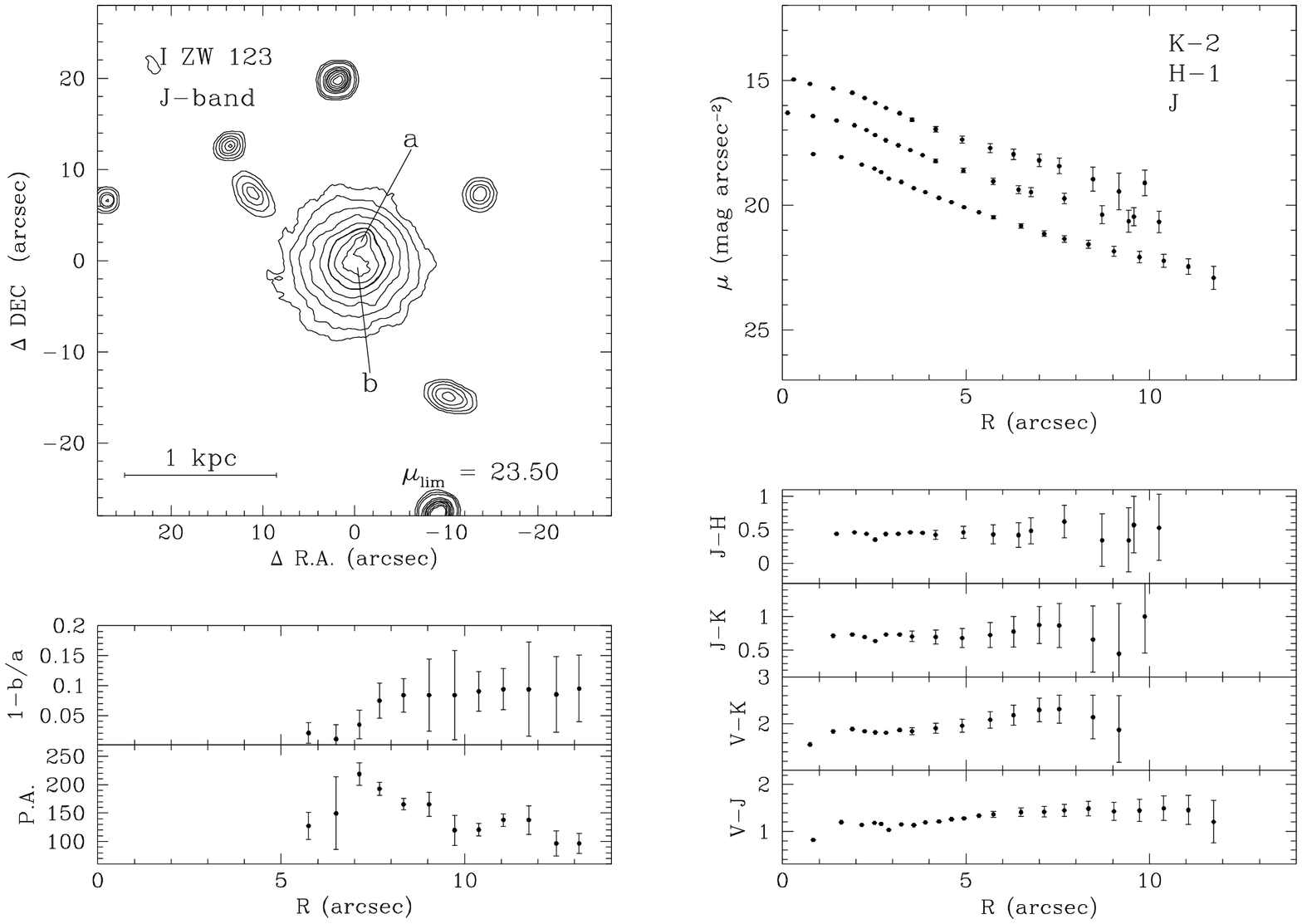}
\caption{\mbox{Continued}}
\label{Fig:SBP8}
\end{figure*}

\addtocounter{figure}{-1}
\begin{figure*}
\includegraphics[angle=270,width=\textwidth]{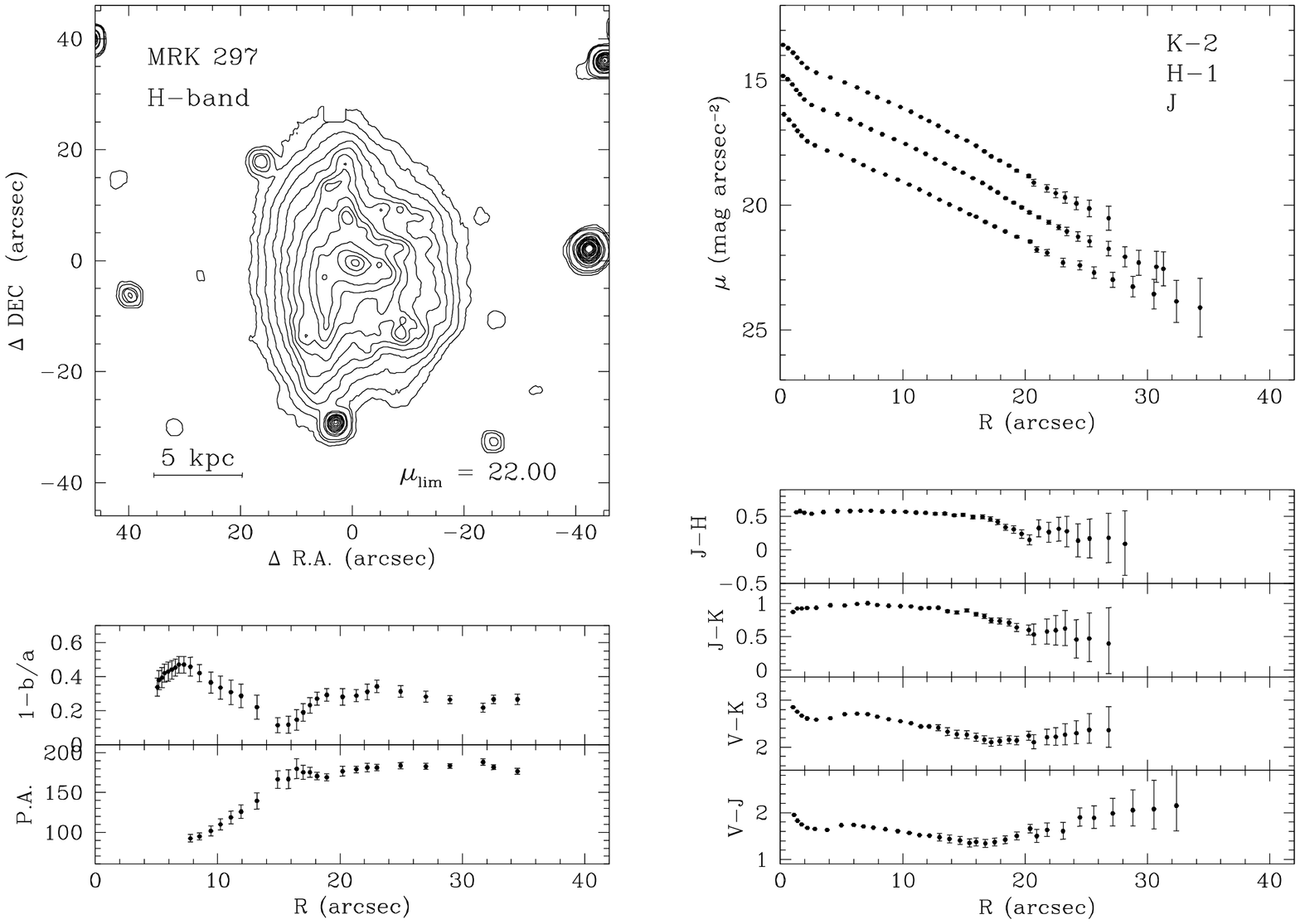}
\caption{Continued}
\label{Fig:SBP9}
\end{figure*}

\clearpage

\begin{figure*}   
\includegraphics[angle=270,width=\textwidth]{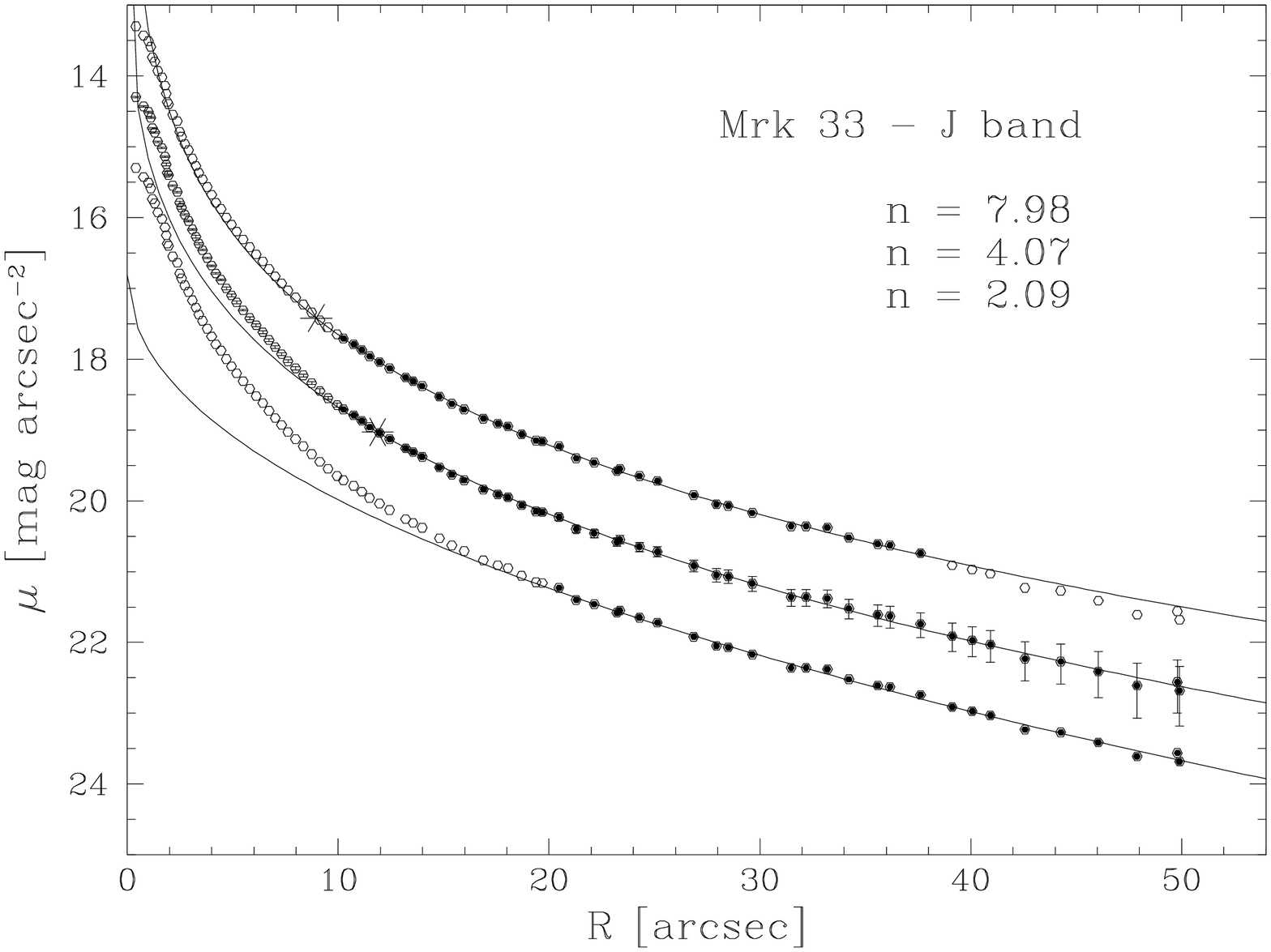}
\caption{Three examples of a S{\'e}rsic fit to the $J$-band light profile of 
Mrk~33 are shown. Filled symbols identify the radial interval over which the
profile was fitted; stars mark the  effective radius and effective surface
brightness. The middle profile is shifted up by 1 mag, the upper profile by 2
mags. Notice how all fits appear to be equally good (rms is $\lesssim 0.04$ 
mags). See text and Table~\ref{Tab:fitmrk335} for further details.}
\label{Fig:mrk35sersic}
\end{figure*}

\begin{figure*}   
\includegraphics[angle=0,width=\textwidth]{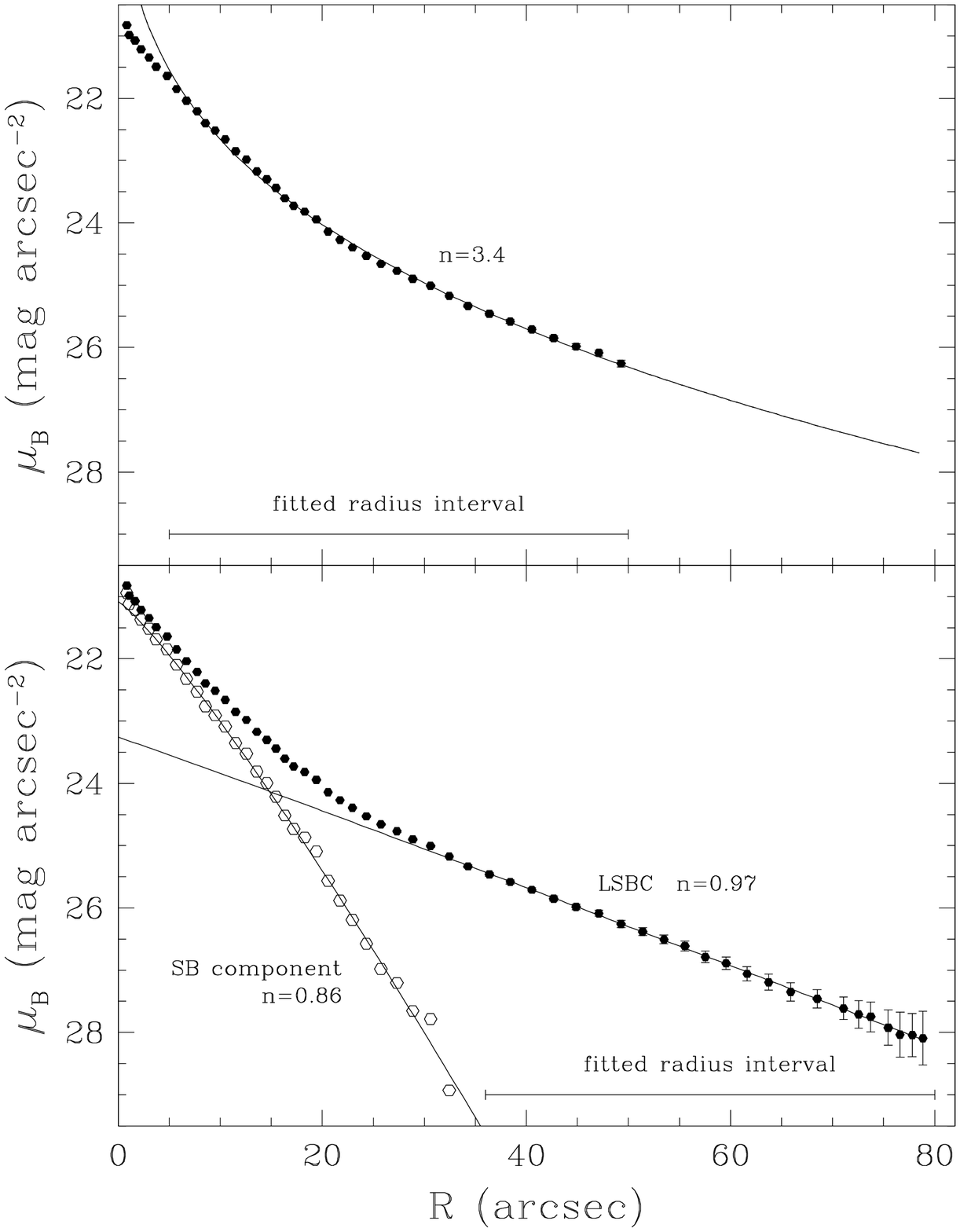}  
\caption{
Upper panel: the $B$-band light profile of VII~Zw~403, plotted down to
$\mu_B \simeq 26.3 \sbb$, shows a relatively smooth curvature, and
can be best fit, in the range $5''<R<50''$, by a S{\'e}rsic law with $n=3.40$.
Lower panel: A deeper light profile, reaching down to $\mu_B \simeq 28.0 \sbb$, 
shows clearly the presence of a LSB component, which can be best fit, in
the interval $36''<R<80''$ by a S{\'e}rsic law with $n=0.97$.
By subtracting the LSB fit to the observed light profile, we obtain the 
starburst profile (empty dots), which turns out to be well described by a
S{\'e}rsic law with $n=0.86$.} 
\label{Fig:VIIZw403}   
\end{figure*}

\begin{figure*}   
\includegraphics[angle=270,width=\textwidth]{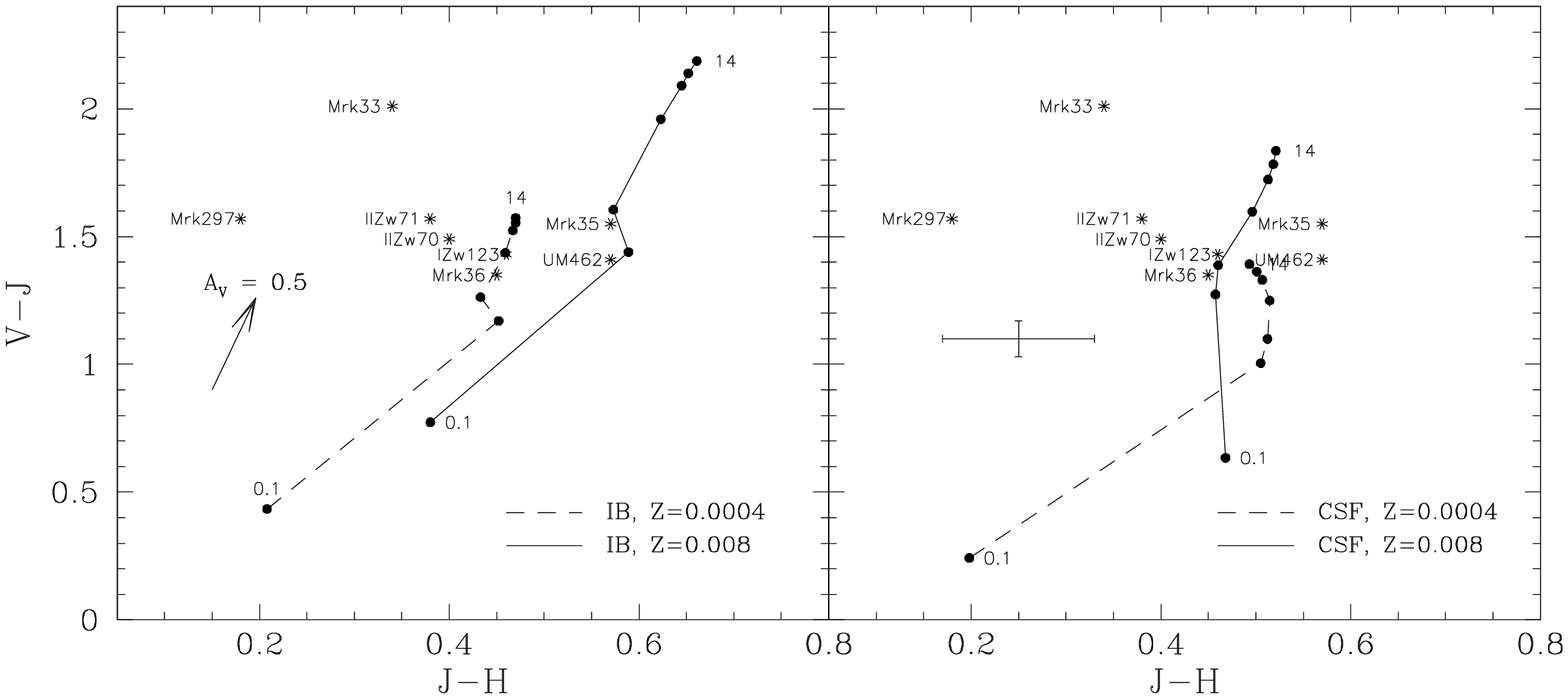}  
\caption{\vj\ vs \jh\ diagram for the LSBC of the sample 
galaxies. We also plot the evolutionary tracks of metallicity $z=0.0004$ and 
$z=0.008$ for an instantaneous burst (IB, left panel) and for continuous star 
formation (CFS, right panel). The arrow in the left panel indicates
the shift in the plot due to an extinction of 0.5 V magnitudes, while the cross
in the right panel shows the typical uncertainty.} 
\label{Fig:modelos}   
\end{figure*}

\begin{figure*}   
\includegraphics[angle=0,width=\textwidth]{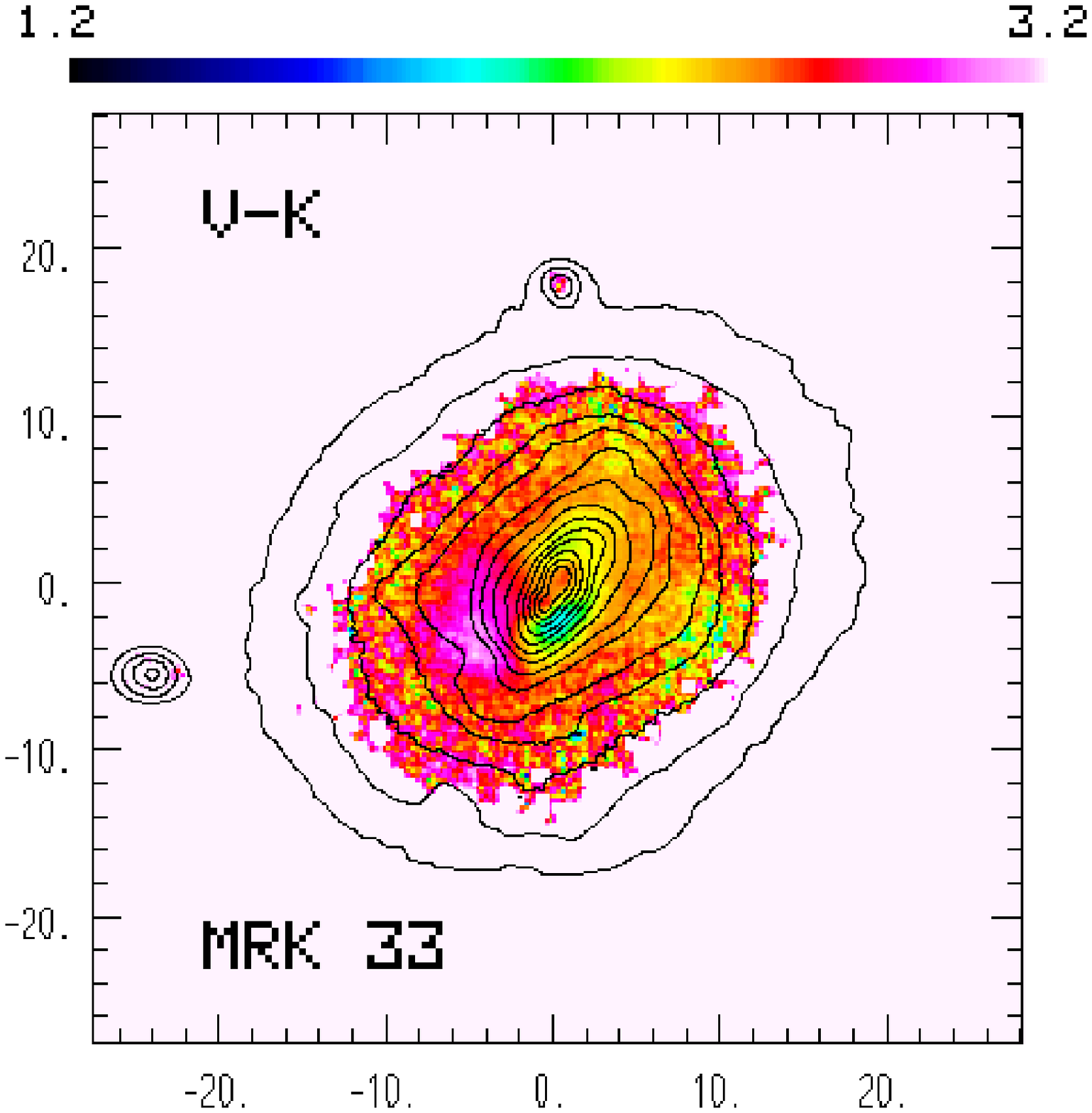}
\caption{\vk\ color map of Mrk~33 overlaid with the $B$-band isophotes.
The linear color scale is shown in the panel. North is to the top and east is 
to the left. Axis units are in arcseconds. The lowest isophote level is 
$\mu_B= 23.5$ mag arcsec$^{-2}$; isophotes are spaced 0.5 mag apart.}
\label{Fig:VKmkn33colormaps}
\end{figure*}

\begin{figure*}   
\includegraphics[angle=0,width=\textwidth]{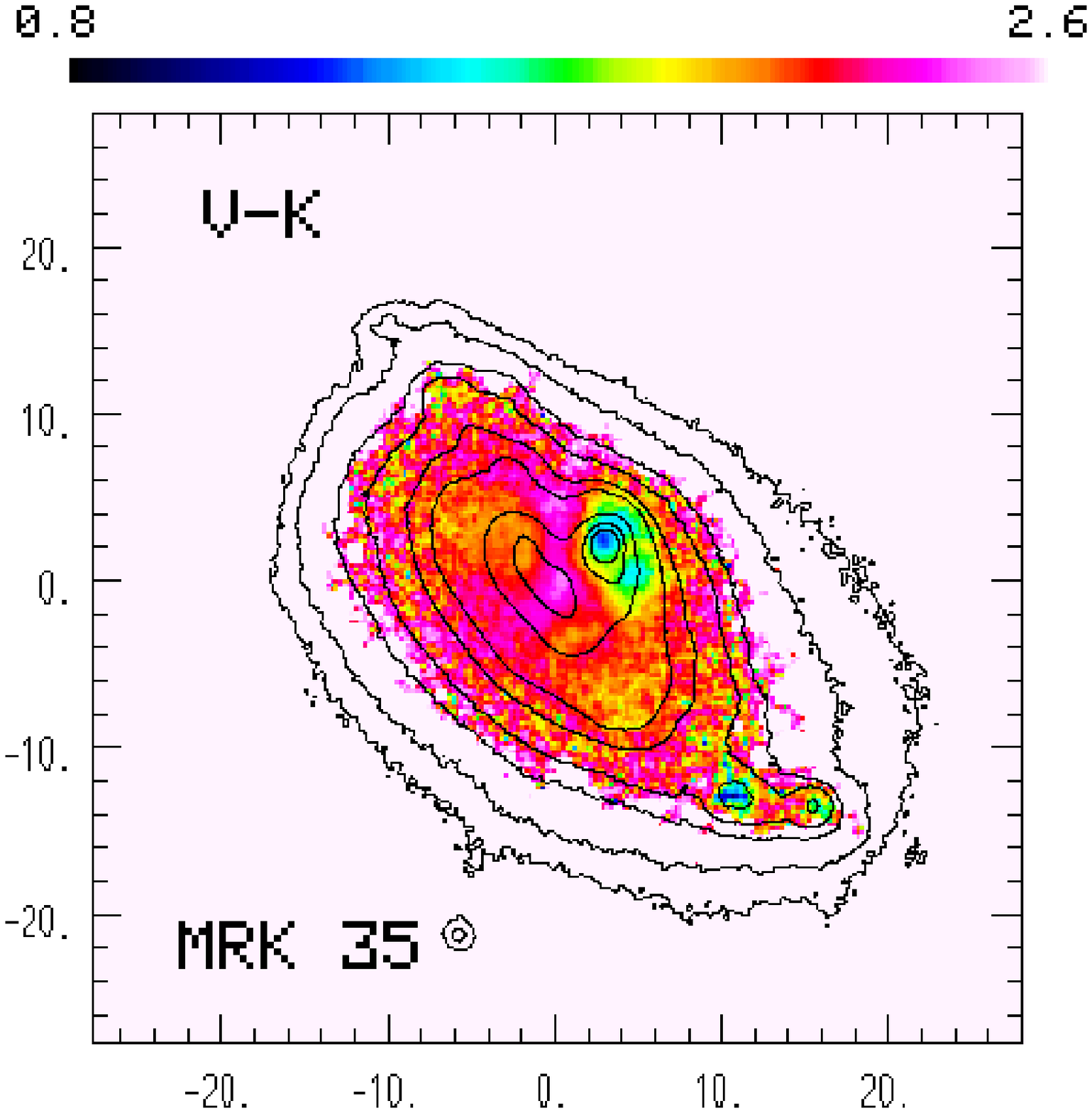}
\caption{\vk\ color map of Mrk~35 overlaid with the $B$-band isophotes.
The linear color scale is shown in the panel. North is to the top and east is 
to the left. Axis units are in arcseconds. The lowest isophote level is 
$\mu_B = 23.0$ mag arcsec$^{-2}$; isophotes are spaced 0.5 mag apart.}
\label{Fig:VKmkn35colormaps}
\end{figure*}

\begin{figure*}   
\includegraphics[angle=0,width=\textwidth]{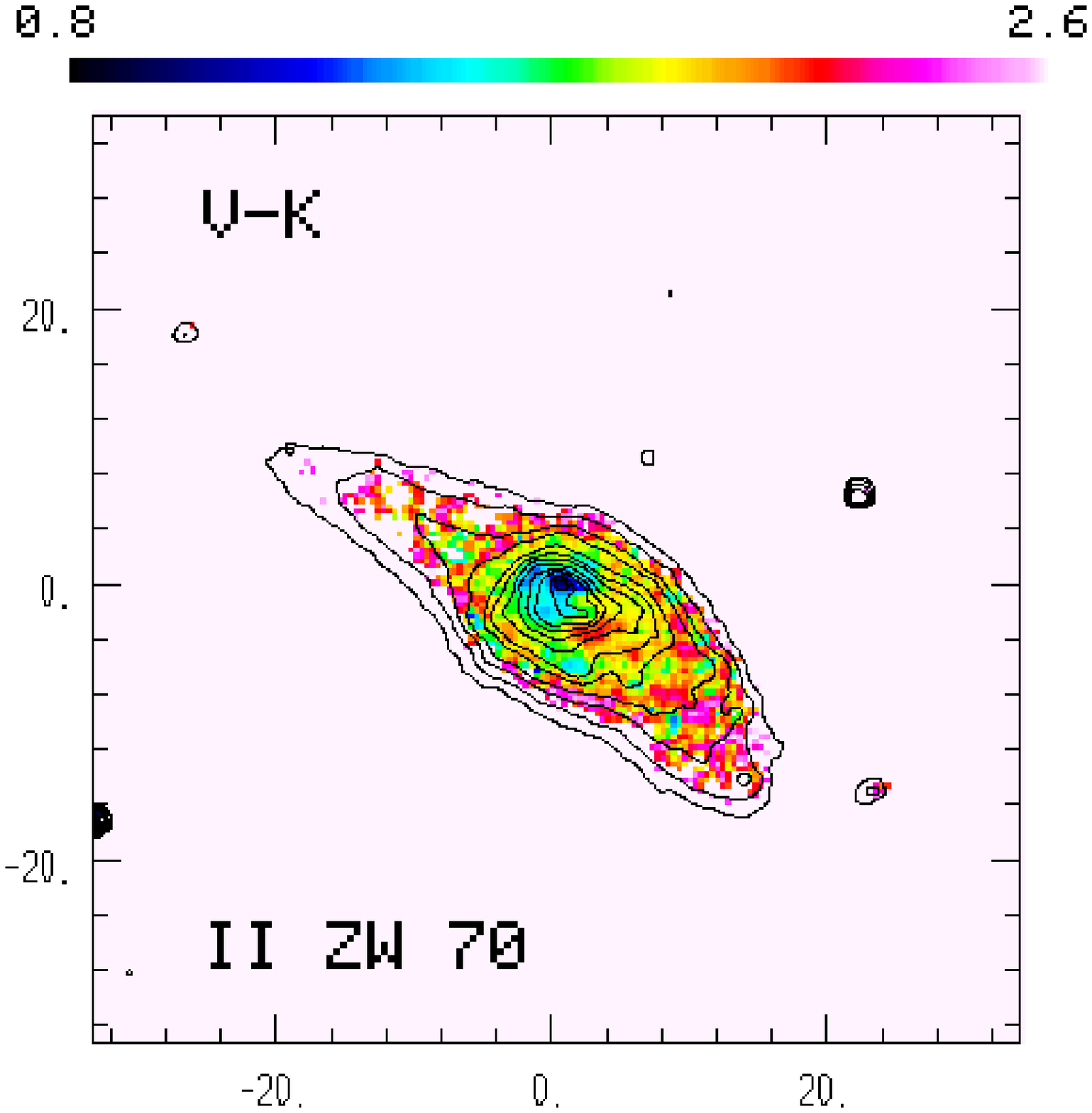} 
\caption{\vk\ color map of II~Zw~70 overlaid with the $V$-band isophotes.
The linear color scale is shown in the panel. North is to the top and east is 
to the left. Axis units are in arcseconds. The lowest isophote level is 
$\mu_V = 23.5$ mag arcsec$^{-2}$; isophotes are spaced 0.5 mag apart.}
\label{Fig:VKiizw70colormaps}
\end{figure*}

\begin{figure*}   
\includegraphics[angle=0,width=\textwidth]{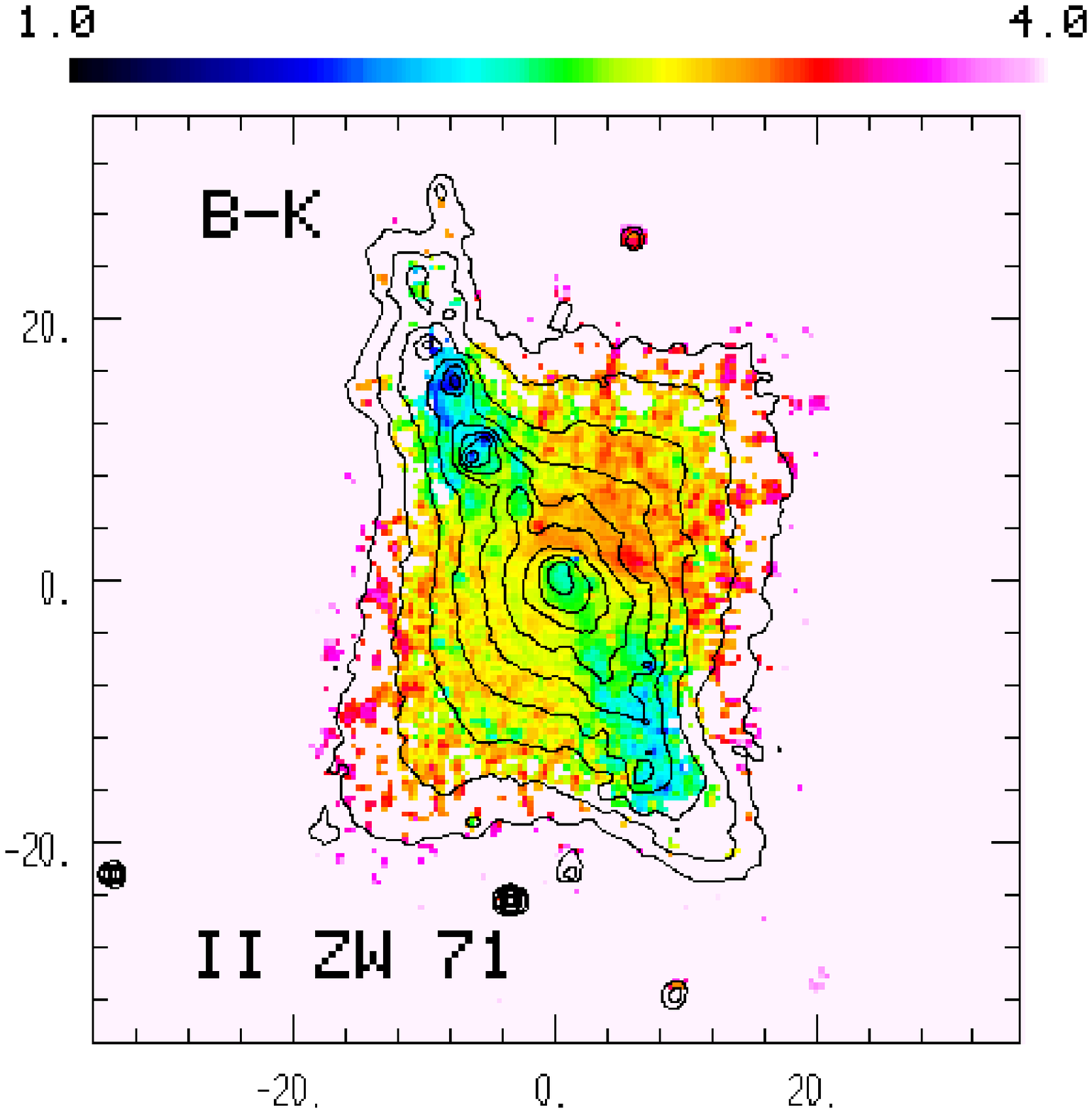}
\caption{\bk\ color map of II~Zw~71 overlaid with the $B$-band isophotes.
The linear color scale is shown in the panel. North is to the top and east is 
to the left. Axis units are in arcseconds. The lowest isophote level is 
$\mu_B = 24.0$ mag arcsec$^{-2}$; isophotes are spaced 0.5 mag apart.}
\label{Fig:VKiizw71colormaps}
\end{figure*}

\begin{figure*}   
\includegraphics[angle=0,width=\textwidth]{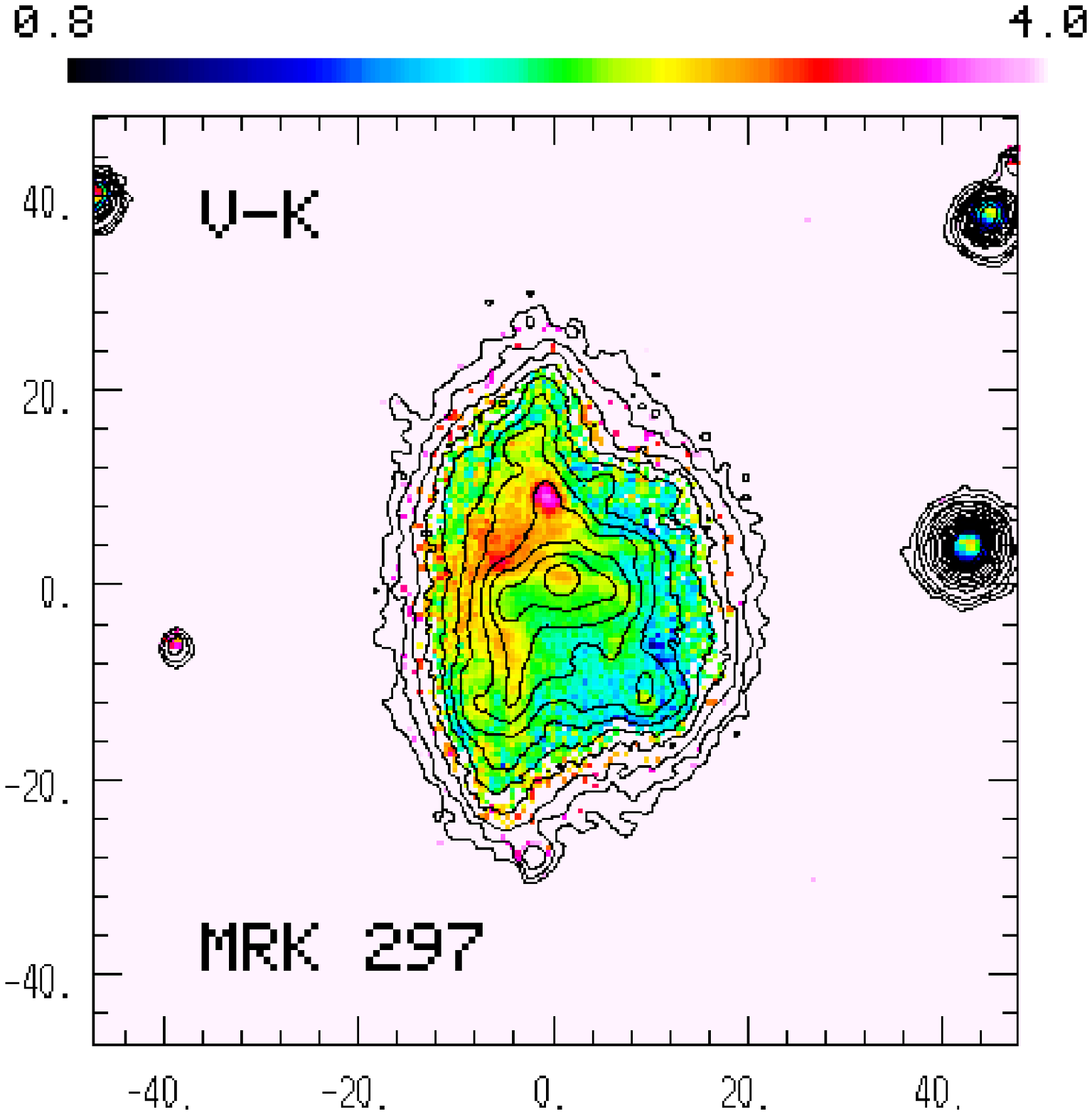}
\caption{\vk\ color map of Mrk~297 overlaid with the $V$-band isophotes.
The linear color scale is shown in the panel. North is to the top and east is 
to the left. Axis units are in arcseconds. The lowest isophote level is 
$\mu_V = 24$ mag arcsec$^{-2}$; isophotes are spaced 0.5 mag apart.}
\label{Fig:VKmkn297colormaps}
\end{figure*}

\end{document}